\documentclass[11pt,twocolumn]{article}
\usepackage[ansinew]{inputenc}
\usepackage[T1]{fontenc}
\usepackage{geometry}
\usepackage{graphicx}
\usepackage{textcomp}

\usepackage{color}
\definecolor{engray}{gray}{.5}
\definecolor{vermelho}{rgb}{.8,0.0,0.0}
\definecolor{verde}{rgb}{.0,0.8,0.0}

\usepackage{natbib}
\usepackage{latexsym,amsmath,amssymb}

\newcommand{\rem}[1]{}

\begin{document}
\title{\vspace{-0.5in}Who Replaces Whom? Local versus Non-local Replacement in Social and Evolutionary Dynamics}
\begin{small}
\author{Sven Banisch$^*$ and Tanya Ara\'{u}jo$^{**}$ \\
* Mathematical Physics, Bielefeld University, Germany \\
\emph{sven.banisch@UniVerseCity.De}\\
** ISEG - Technical University of Lisbon (TULisbon) and\\
Research Unit on Complexity in Economics (UECE), Portugal\\
\emph{tanya@iseg.utl.pt}} \end{small}
\date{}
\maketitle

\vskip 2truecm

\begin{abstract}
\small
In this paper, we inspect well--known population genetics and social dynamics models.
In these models, interacting individuals, while participating in a self-organizing process, give rise to the emergence of complex behaviors and patterns.
While one main focus in population genetics is on the adaptive behavior of a population, social dynamics is more often concerned with the splitting of a connected array of individuals into a state of global polarization, that is, the emergence of speciation.
Applying computational and mathematical tools we show that the way the mechanisms of selection, interaction and replacement are constrained and combined in the modeling have an important bearing on both adaptation and the emergence of speciation.
Differently (un)constraining the mechanism of individual replacement provides the conditions required for either speciation or adaptation,
since these features appear as two opposing phenomena, not achieved by one and the same model.
Even though natural selection, operating as an external, environmental mechanism, is neither necessary nor sufficient for the creation of speciation, our modeling exercises highlight the important role played by natural selection in the interplay of the evolutionary and the self--organization modeling methodologies.

{\it Keywords}:
{\bf Emergence},{\bf Self-organization},{\bf Agent Based Models}, {\bf Speciation}, {\bf Markov chains}.

{\it MSC:} 37L60, 37N25, 05C69.
\end{abstract}

\section{Introduction}

There are two important phenomena observed in evolutionary dynamical systems of any kind: \emph{Self-organization} and \emph{Emergence}.
Both phenomena are the exclusive result of endogenous interactions of the individual elements of an evolutionary dynamical system.
Emergence characterizes the patterns that are situated at a higher macro level and that arise from interactions taking place at
the lower micro level of the system. Self-organization, besides departing from the individual micro interactions,
implies an increase in order of the system, being usually associated to the promotion of a specific functionality and to the generation of patterns.
Typically, complex patterns emerge in a system of interacting individuals that participate in a self-organizing process.
Self-organization is more frequently related to the process itself, while emergence is usually associated to an outcome of the process.

Although less frequently mentioned, the emergence of patterns from self-organizing processes may be strongly dependent on \emph{Locality}.
Emergence and self-organization are not enough to distinguish between two important and quite different circumstances: the presence of an influence that impacts the system globally and, conversely, the absence of any global influence and the lack of information about any global property of the system. In the latter case, the system itself is the exclusive result of local interactions.

Such a global influence (entity or property) is often associated with the concept of \emph{Environment}. Noteworthy, the latter circumstance may be considered a case of the former: when that global entity does not exist, the environment for each agent is just the set of all the other agents. Conversely, when the global entity exists, it is considered part of the environment and may have an inhomogeneous impact on the individual dynamics.

Regardless of the environmental type, economical, ecological and social environments share as a common feature the fact that the agents operating in
these environments usually try to improve some kind of utility, related either to profit, to food, to reproduction or to comfort and power.
A general concept that is attached to this improvement attempt is the idea of \textit{Adaptation}.

In the economy, adaptation may be concerned with the development of new products to capture a higher market share or with the improvement of the production processes to increase profits: that is, innovation. In ecology, adaptation concerns better ways to achieve security or food intake or reproduction chance and, in the social context, some of the above economical and biological drives plus a few other less survival-oriented needs. In all cases, adaptation aims at finding strategies to better deal with the surrounding environment (\cite{Araujo2009}).


Natural selection through fitness landscapes or geographic barriers are good examples how global influences are considered when modeling adaptation in an
evolutionary process. On the other hand, adaptation also operates in many structure generating mechanisms that can be found in both physical and social sciences but that are built on the exclusive occurrence of local interactions.

\begin{figure}[h]
\centering
\includegraphics[width=0.7\linewidth]{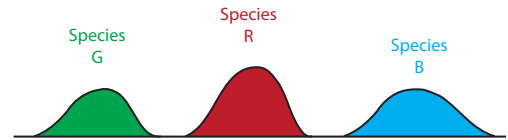}
\caption{Schematic illustration of organic diversity.}
\label{fig:Distribution01}
\end{figure}

In biology, the ultimate domain of evolution and natural selection, we are confronted with  tremendous organic diversity -- virtually infinite forms and shapes none of which found twice -- but the distribution is well--structured in a way that allows us to order this diversity and to speak of species, families, orders etc.
A quite illustrative description is given by the evolutionary geneticist Theodusius Dobzhanski (\cite{Dobzhanski1970}: p.21):
\begin{quote}
Suppose that we make a fairly large collection, say some 10,000 specimens, of birds or butterflies or flowering plants in a small territory, perhaps 100 square kilometers.
No two individuals will be exactly alike.
Let us, however, consider the entire collection.
The variations that we find in size, in color, or in other traits among our specimens do not form continuous distributions.
Instead, arrays of discrete distributions are found.
The distributions are separated by gaps, that is, by the absence of specimens with intermediate characteristics.
We soon learn to distinguish the arrays of specimens to which the vernacular names English sparrow, chickadee, bluejay, blackbird, cardinal, and the like, are applied.
\end{quote}
If we had to make a visual representation of this description of intra-- and interspecies variations it would perhaps look like the multi-modal distribution shown in Figure \ref{fig:Distribution01}.
What we call a species, is in fact some norm or mean characteristics of a cluster of individuals.

\begin{figure}[h]
\centering
\includegraphics[width=.5\linewidth]{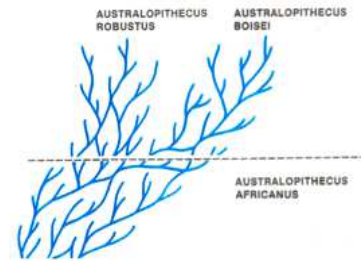}
\caption{Illustration of a speciation event. We are grateful to Andreas Dress for providing us this figure.}
\label{fig:Speciation}
\end{figure}

Evolutionary theory is ultimately a theory about the history which led to such a pattern.
And if the organic diversity we observe nowadays evolved in a way that is characterized by some kind of >>Tree of Live<<, then there must be events that may lead to the split of a connected set of individuals (protospecies) into (at least) two sets that are not connected any longer (see Figure \ref{fig:Speciation}).
In biology, this is called \emph{Speciation}.
As we will see in this article, though, the generation of such a split with simple but well--known evolutionary models in which "natural selection impels and directs evolutionary changes" (ibid. p.2) is not straightforward.
It so happens that constraints on the interaction behavior are required.

The phenotype of living beings is not the only domain where patterns of structured diversity as illustrated in Figure \ref{fig:Distribution01} are observed.
Phenomena include certain phases of structure formation in physical cosmology, distribution of cultural behavior, languages and dialects, herd behavior in finance, among others.


Especially for the latter examples in the field of socio-cultural dynamics a variety of models has been proposed which do not rely on the evolutionary concept of (natural) selection.\footnote{\cite{Castellano2009} provide a comprehensive overview over models in this field.}

They are rather based on the idea of exclusively \emph{Local Interactions (LI)} implemented in form of a system of agents that interact locally according to simple rules like assimilation or conformity. In these systems, finding strategies to better deal with the surrounding environment (and thus improving fitness) is not constrained by any global property.
It may, however, be constrained by local (individual) rules.

As we shall see later in this paper, constraints on the mechanisms of selection, interaction and replacement and the way they are combined in the modeling of an evolutionary process have an important bearing on both adaptation and emergence of speciation. Locality operating in each of these mechanisms seems to be the fundamental modeling principle by which emergence of a multi-modal distribution as shown in Figure  \ref{fig:Distribution01} can be explained.
On the basis of these observations about the >>modelability<< of speciation  with evolutionary and self-organisatory models, we study in this paper the conditions and mechanisms required for speciation and the emergence of a multi-modal distribution.

In this analysis, we use computational (Section \ref{sec:computation}) as well as mathematical (Section \ref{sec:mathematic}) arguments.
Our models simulate how a population of individuals evolves in time in an abstract attribute space $({\bf S})$ that represent phenetic traits, attitudes, verbal behavior, etcetera.
Modeling agents as points in an attribute space of this kind is of course a highly artificial abstraction from the complexity and multi--dimensionality of real agents.


For the purposes of this paper, let us conceptualize an \emph{Interaction Event}, defining the system evolution from one time step to the other, by the following three components:
\begin{enumerate}
\item
selection of agents,
\item
application of interaction rules,
\item
replacement of agents.
\end{enumerate}
Any interaction event (e.g., mating, communication,...) that takes place in the course of a simulation of the model consists of the sequential application of these three steps.
The reason to dissect the interaction events in this way is two--fold:
\begin{enumerate}
\item
we want to look at the dynamical and structural effects of constraints applied to each of the three components independently;
\item
the scheduling of interaction events may have a crucial effect on the model behavior, and
with the distinction between selection and interaction on the one hand, and replacement on the other, we are able to make this effect explicit.

\end{enumerate}

The way interaction events are scheduled in the implementation of the models is not always given much importance in existing simulation studies.
In the presence of constraints on the selection and interaction mechanisms, however, the outcome as well as the dynamical properties depend in a crucial way on the different choices.
On the other hand, there are studies that do analyze the differences between synchronous and asynchronous update (see, for instance, \cite{Hubermann1993,Banisch2010})
as well as studies on non--overlapping (NOLG) and respectively overlapping generations (OLG) in biology and economics (for instance, \cite{Kehoe1984}).


Here we show that especially when the interaction is constrained (as in the case of assortative mating) there emerges an important qualitative difference between OLG and NOLG models.
Namely, speciation is observed in the former, but not in the latter case, whereas adaptation is favored by the latter and hindered by the former.
However, by the distinction of selection, interaction and replacement we are able to show that in fact the difference between local and non-local replacement plays the determinant role (and not the distinction between OLG and NOLG).
Even though locality also impacts selection and interaction mechanisms, it is on the replacement mode where relies the fundamental difference with respect to the conditions required for either adaptiveness or speciation. 

\rem{

\emph{The sucessive layers in the following scheme helps to summarize our main research concern\footnote{To be completed in the Discussion (Sect.4, Table 1).}.}

\begin{center}
\emph{Complex Systems} \\
\bigskip
\emph{Self-organizing Process} $ \Rightarrow $ \emph{Emergence of Patterns}\\
\bigskip
\emph{Constraints on the SO Process } $ \Rightarrow $ \emph{Emergence of Specific Patterns}\\
\bigskip
\emph{Constraints on Replacement}  $ \Rightarrow $ \emph{Emergence of Speciation}\\
\end{center}

\bigskip

\bigskip
}

This paper is organized as follows: Section 2 addresses the main issues of both the fitness landscape and the self-organizing models from a computer simulation framework. In both cases, microscopic implementation rules are tested against their capability of  reproducing adaptiveness and speciation. In Section 3, the emergence of speciation is analytically shown to be dependent on the choice of different replacement modes. This is accomplished through a probabilistic description of a minimal model of just three phenetic traits where the transition probabilities between traits follow a Markov chain. Section 4 is targeted at presenting concluding remarks and a framework that relates interaction events to the emergence of collective structures in adaptive and self-organizing complex systems.

\bigskip

\bigskip

\section{From Adaptive Dynamics to Cluster Formation}
\label{sec:computation}

\subsection{Adaptive Walks on Fitness Landscapes}

In biology, and population genetics in particular, adaptive walks on fitness landscapes have been studied intensively.
The main questions addressed by fitness landscapes approaches are related to the possible structure of the landscapes (e.g., \cite{Kauffman1993}), to how populations climb an adaptive peak in the landscape (e.g., \cite{Fisher1958}), and to the circumstances under which a population might wander from one peak to another by crossing adaptive valleys (e.g. \cite{Wright1932}).


One of the best--known models for populations on fitness landscapes is the Wright-Fisher model with non--overlapping generations (sometimes called Wright-Fisher sampling and shortened in the sequel by WF model, see \cite{Crow1970} and also \cite{Drossel2001}).
Consider a population of $N$ individuals which is said to constitute the original generation ($g = 0$).
We consider only the case of sexual reproduction in this paper, in which the genotype of a new--born individual is obtained by the recombination of the genoms of two randomly chosen parent individuals.
As noted above, the choice of two parents and the application of a recombination rule is referred to as interaction (or mating) event.
In the WF model, $N$ such mating events are performed until a new generation of $N$ individuals is complete.
As soon as it is complete, the parent generation is canceled and the process is repeated taking the new generation as parents.
Therefore, in the WF model the population size is always maintained at $N$.
We will denote the generation number by $g = 0,1,2,\ldots$.

We implemented this simple model and performed simulations on different toy fitness landscapes.
The microscopic rules involved into the creation of a new individual, that is, the mating event, are as follows:
\begin{enumerate}
\item
selection of two individuals with a probability proportional to their fitness,
\item
application of recombination and mutation rules,
\item
replacement of an agent from the parent generation.
\end{enumerate}
In this toy model, we consider only one phenetic trait (locus) that takes discrete values (from 0 to 99).
We denote the traits of the two chosen parent individuals $i$ and $j$ as $x_i$ and $x_j$ respectively and model recombination by taking the average of the two, $x_{new} = (x_i+x_j)/2$.
To model mutations we add a random value to $x_{new}$.
In the WF model, $x_{new}$ is stored at an arbitrary place in the children array and one of the main objectives of this paper is clarify that this has important consequences for the model dynamics.

\begin{figure*}[htp]
\centering
\begin{tabular}{ccc}
\includegraphics[width=0.28\linewidth]{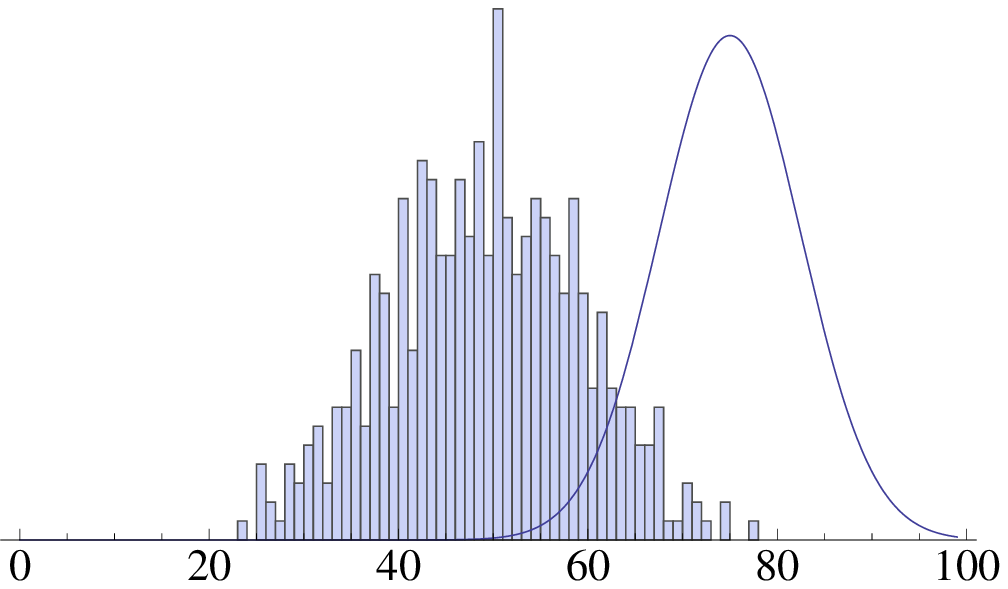}&
\includegraphics[width=0.28\linewidth]{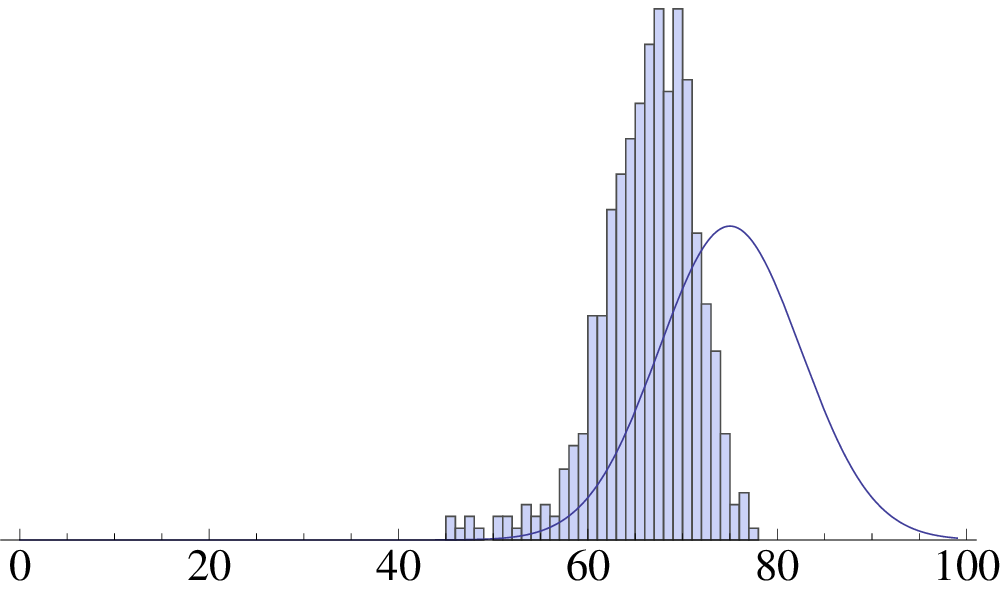}&
\includegraphics[width=0.28\linewidth]{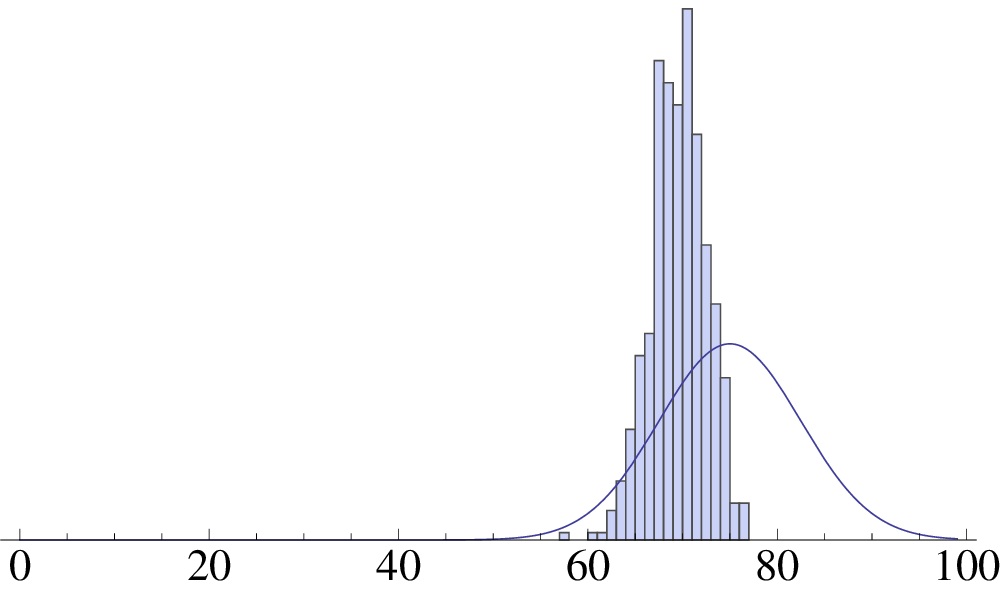}\\
\includegraphics[width=0.28\linewidth]{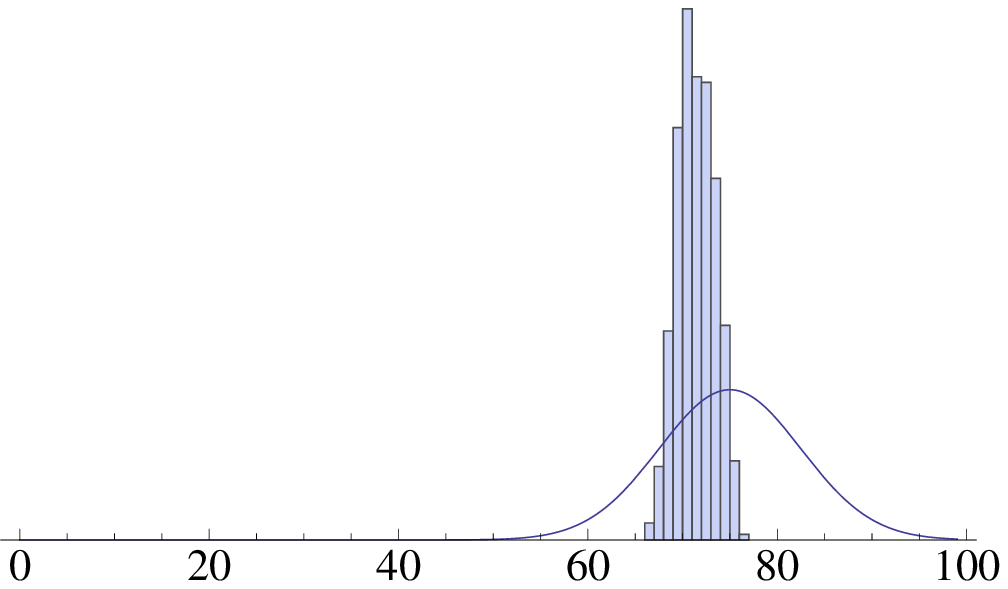}&
\includegraphics[width=0.28\linewidth]{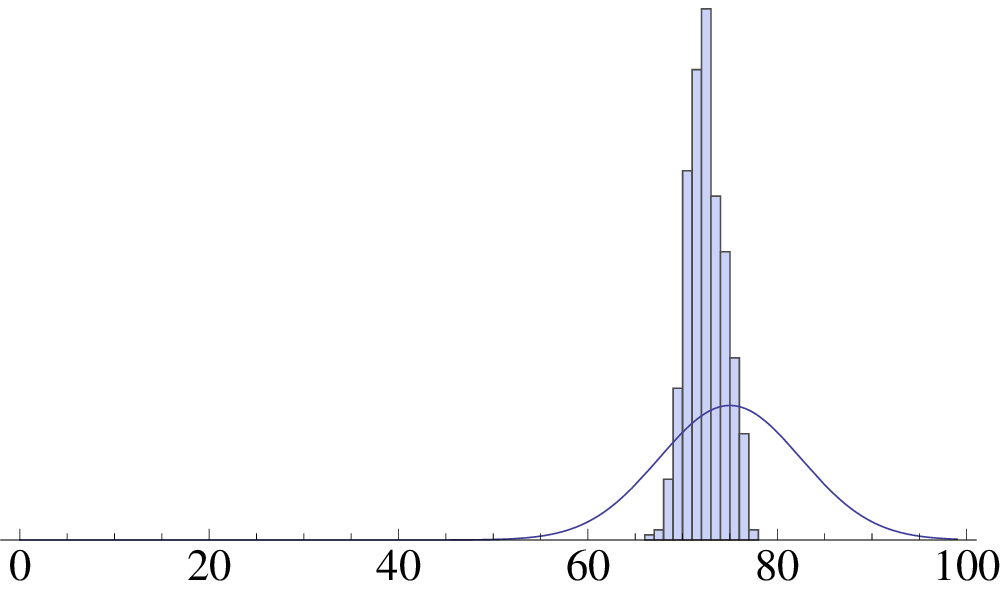}&
\includegraphics[width=0.28\linewidth]{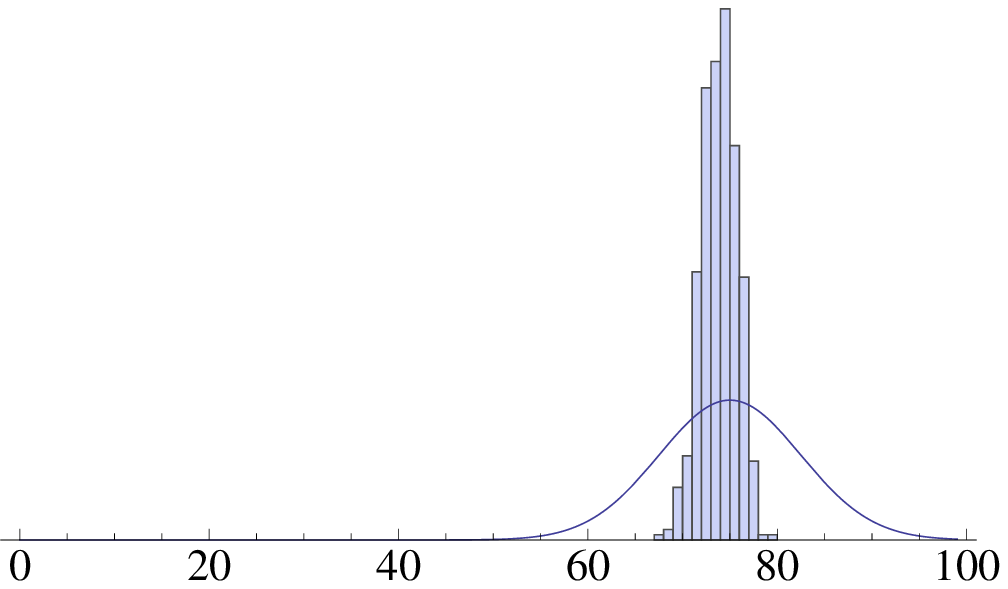}
\end{tabular}
\caption{WF model approaches an adaptive peak. In the upper row the initial generation ($g=0$) and the distribution after the first ($g=1$) and the second ($g=2$) iteration are shown from left to right. Bottom row shows, from left to right, the 5th, 10th and 20th generation ($g = 5,10,20$).}
\label{fig:OnePeak.WF}
\end{figure*}

An adaptive landscape is introduced into the model by assigning a fitness value to each of the 100 traits.
For the first analysis shown in Figure \ref{fig:OnePeak.WF}, a single--peaked fitness function with a peak at trait 75 is used and the fitness assigned to trait $x$ is given by
\begin{equation}
F(x) = \frac{1}{15} e^{-\frac{2}{225} (-75+x)^2} \sqrt{\frac{2}{\pi }} = \mathrm{N}(\mu,\sigma^2).
\end{equation}
We have used the normal distribution with $\mu= 75$ and $\sigma^2 = 7.5$ in the construction of the fitness landscape (solid line in Figure \ref{fig:OnePeak.WF}).
In the iteration process, individuals are chosen as parents with a probability proportional to $F(x)$, $x$ being the trait of the respective individual.

For the illustrative model realizations in this section, we set $N = 500$.
Initially, the 500 individuals are distributed in this space according to a normal distribution with mean $\mu= 50$ and $\sigma^2 = 10$ (see first image of Figure \ref{fig:OnePeak.WF}).

This section is mainly thought as an illustration of the different behaviors and patterns generated by certain constraints on the interaction mechanism.
As the qualitative effects of different assumptions become evident and comprehensible in single simulations of the model, there is no need for a rigorous statistical analysis of suites of simulations with varying initial conditions.
Moreover, a mathematical analysis of the model dynamics is presented in the second part of this paper (Section \ref{sec:mathematic}).

\begin{figure*}[htp]
\centering
\begin{tabular}{ccc}
\includegraphics[width=0.28\linewidth]{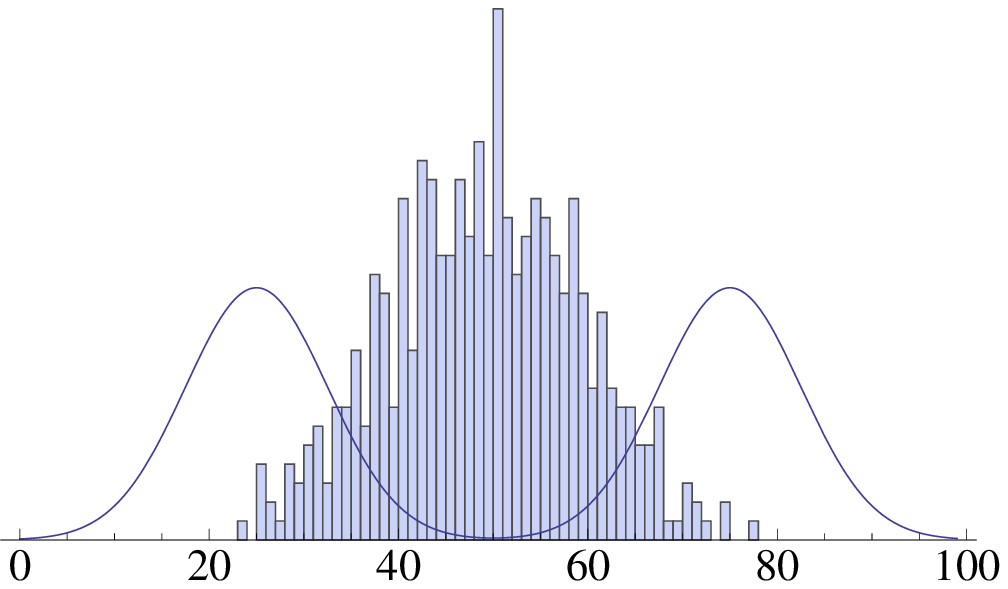}&
\includegraphics[width=0.28\linewidth]{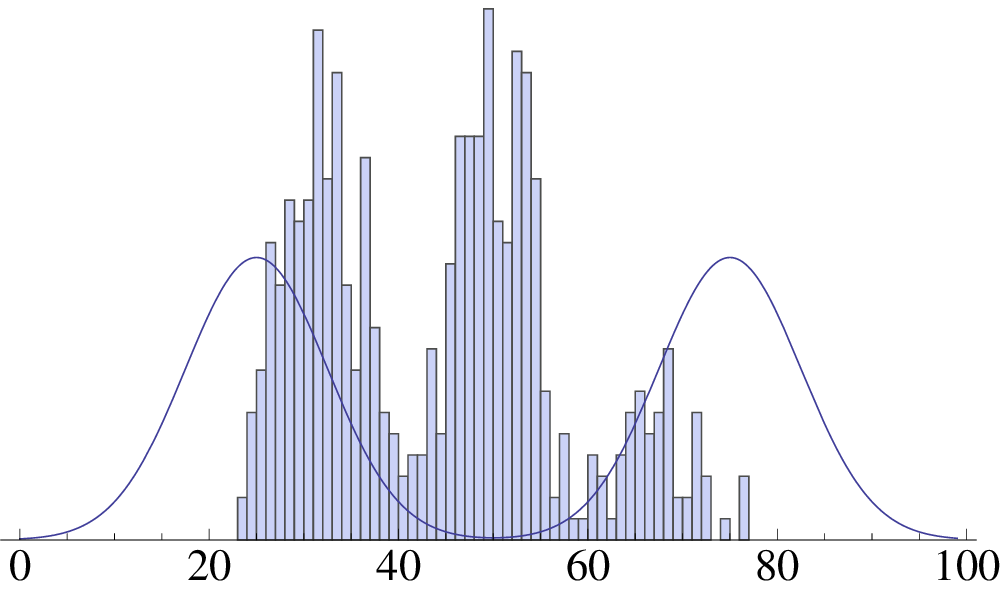}&
\includegraphics[width=0.28\linewidth]{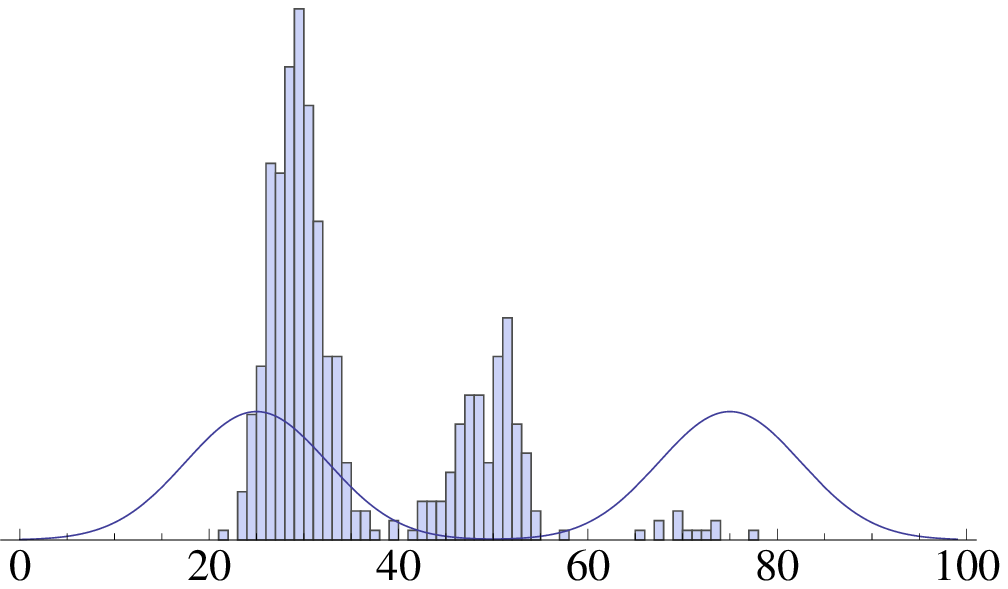}\\
\includegraphics[width=0.28\linewidth]{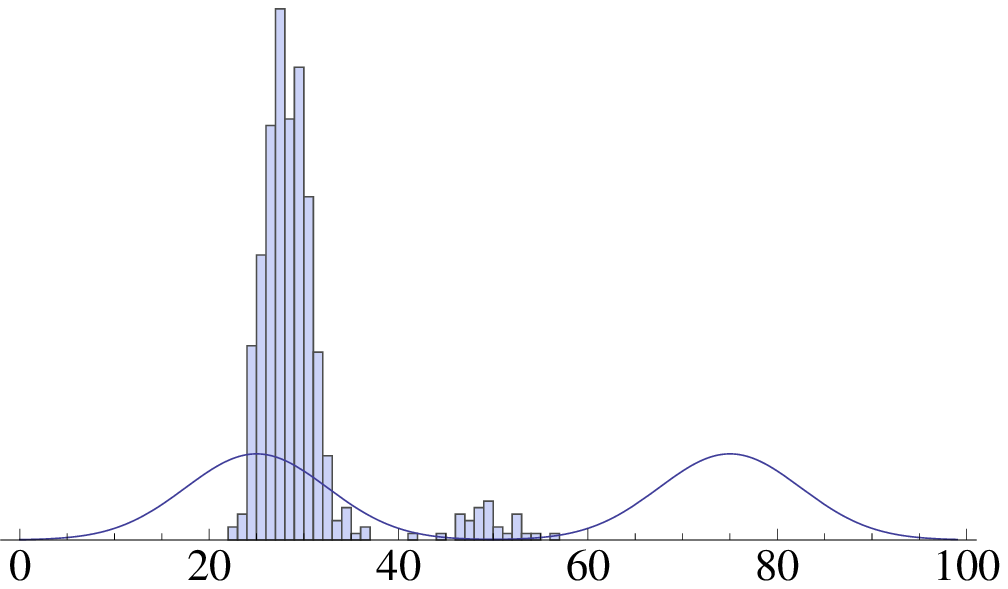}&
\includegraphics[width=0.28\linewidth]{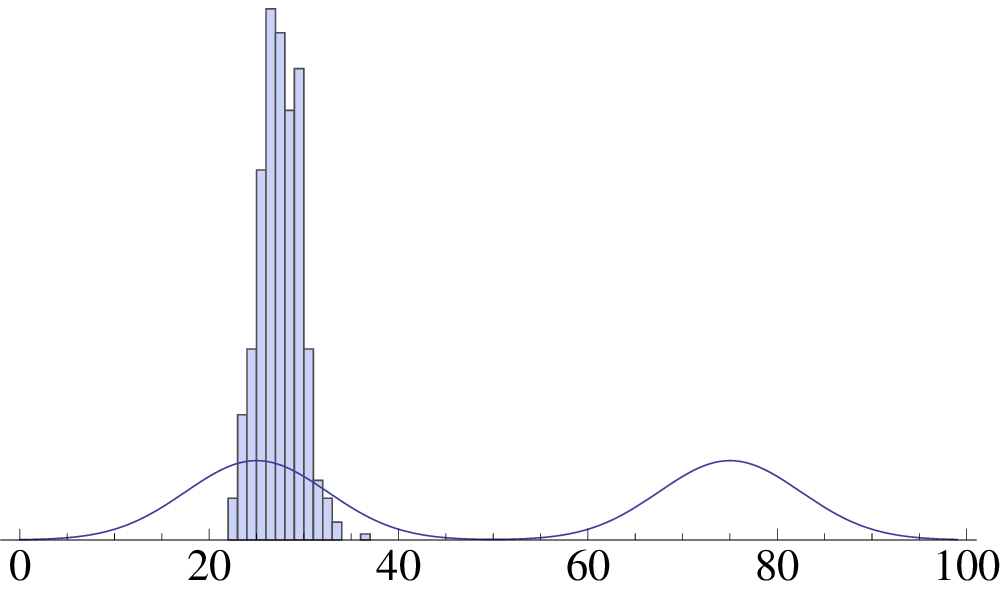}&
\includegraphics[width=0.28\linewidth]{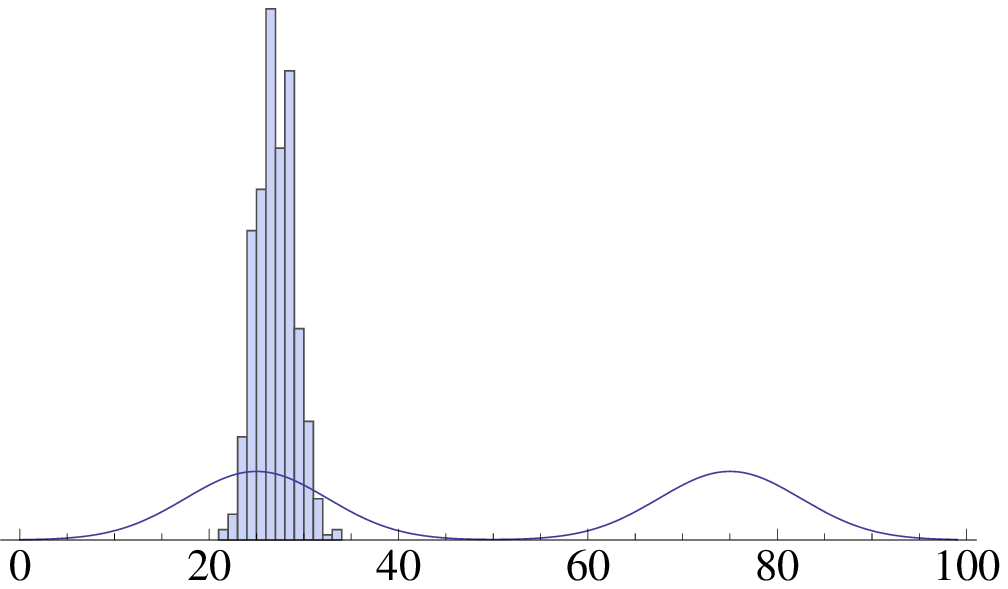}
\end{tabular}
\caption{WF model with a two--peaked fitness landscape approaches a single adaptive peak. From top left to bottom right the initial state ($g=0$) and the first five generations are shown ($g = 1,2,3,4,5$).}
\label{fig:TwoPeaks.WF}
\end{figure*}

\begin{figure*}[htp]
\centering
\begin{tabular}{ccc}
\includegraphics[width=0.28\linewidth]{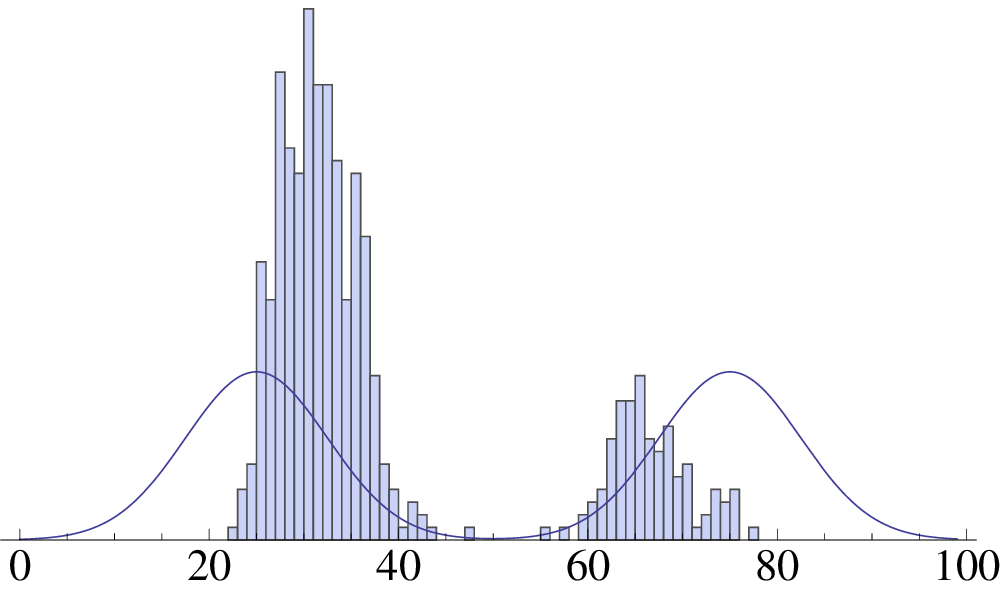}&
\includegraphics[width=0.28\linewidth]{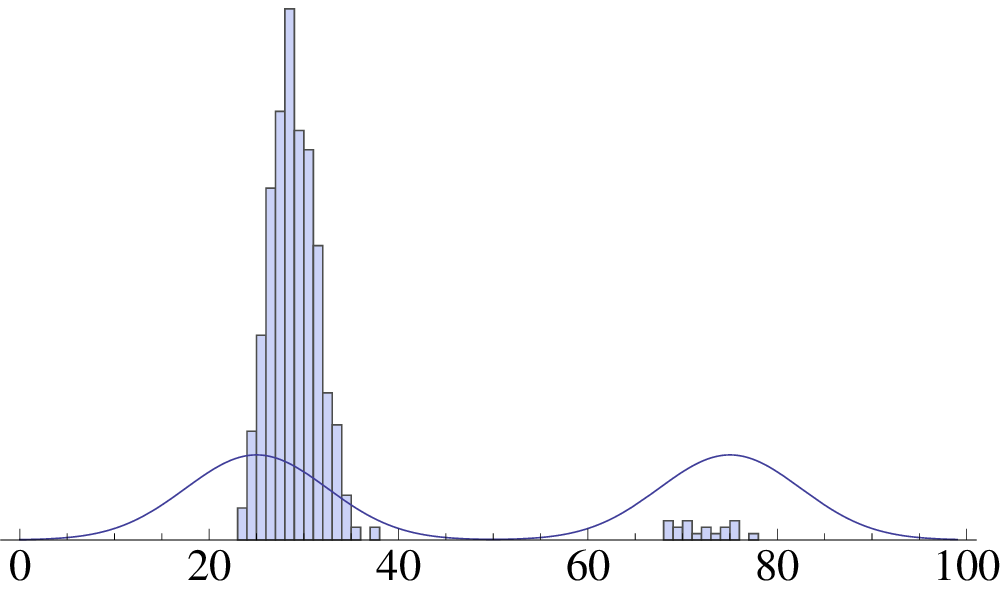}&
\includegraphics[width=0.28\linewidth]{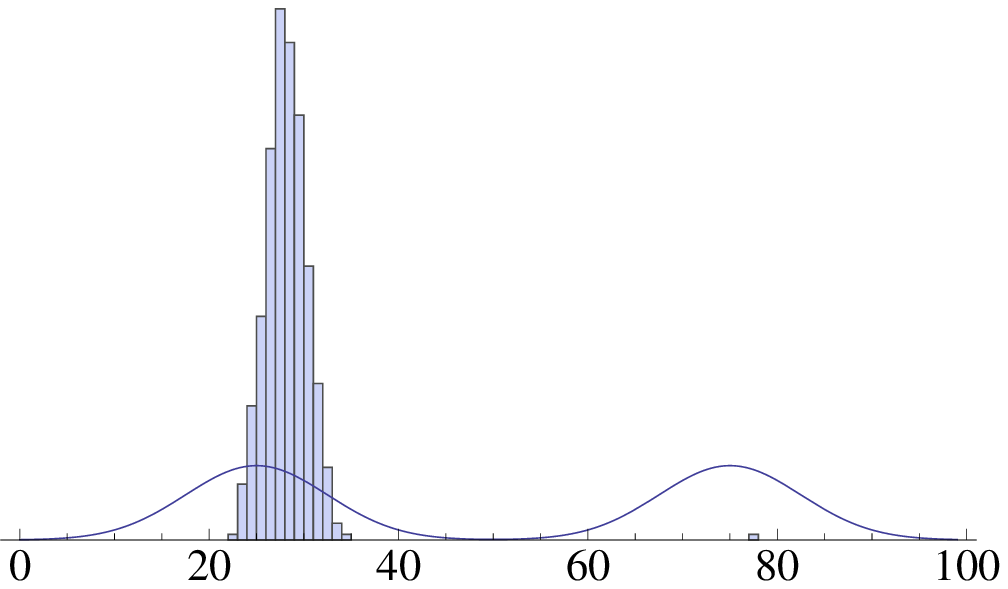}
\end{tabular}
\caption{WF model with a two--peaked fitness landscape and assortative mating approaches a single adaptive peak. From left to right the first three generations are shown ($g = 1,2,3$). The initial state is as in Figure  \ref{fig:TwoPeaks.WF}.}
\label{fig:TwoPeaks.WF.AM10}
\end{figure*}

Figure \ref{fig:OnePeak.WF} shows the first few iterations of the WF model.
The adaptive peak at around 75 is reached within only a few iterations.
Due to mutations, the population does not become fixed at one specific trait, but maintains a certain amount of variation.
Populations simulated with the WF model are very fast in reaching an adaptive peak in the fitness landscape.

\subsection{Sympatric Speciation}

Figure  \ref{fig:OnePeak.WF} shows that the WF model is well--suited to show how a finite population approaches a peak in the fitness landscape.
However, what about speciation?
To get a first insight about whether the splitting of the unimodal initial distribution into a bimodal distribution with two clusters is possible we simulated the model with a two--peaked fitness landscape.
So the difference with respect to the previous simulation is that the fitness function (solid line) has two adaptive peaks, one centered at 25 and the other at 75.
The fitness (that is, the probability of choosing an individual in state $x$) is defined by a mixture of two normal distributions $\mathrm{N}(25,7.5)$ and $\mathrm{N}(75,7.5)$:
\begin{equation}
F(x) = \frac{e^{-\frac{2}{225} (-75+x)^2}}{15 \sqrt{2 \pi }}+\frac{e^{-\frac{2}{225} (-25+x)^2}}{15 \sqrt{2 \pi }}.
\end{equation}

The first five iterations of that model are shown in Figure  \ref{fig:TwoPeaks.WF}.
The initial distribution is as in the previous example.
We see that multi--modal shapes emerge only in the very first few generations of the model.
Namely, after the first and the second iteration, there are three clusters: two located at the peaks and a third one with low fitness in between the other two.
The latter can be seen as hybrid individuals with strong selective disadvantages that are obtained by a recombination of individuals from the different peaks.
However, the disappearance of clustering is very fast and after only four iterations all the population concentrates at one of the peaks.
Hence, in the model it is difficult to generate a stable co--existence of species.


The case considered here is the case of speciation in sympatry: no geographic constraints are assumed to divide the population into reproductive islands or to constrain the mating chances of pairs of individuals in any other way.
A possible explanation why the simulation of sympatric speciation is not possible in the WF model as described above is provided in the seminal paper on sympatric speciation by \cite{Smith1966}.
Smith showed that besides selective forces, it is necessary that the population sizes of the (two) sub--populations are regulated independently.
Because the total population size is usually constant in the WF model (in our case $N = 500$), it does not implement an independent regulation of sub--populations.

An issue frequently discussed in the context of sympatric speciation is assortative mating (see, for instance, \cite{Kondrashov1998,Dieckmann1999} and references therein).
We also simulated the WF model with the additional constraint that two individuals need to be similar in order to produce offsprings.
Two chosen individuals $i$ and $j$ only produce an individual for the new generation if the their difference is small (here $|x_i-x_j| < 10$).
The microscopic rules become:
\begin{enumerate}
\item
selection of two individuals with a probability \emph{proportional to their fitness},
\item
application of recombination and mutation rules \emph{if the individuals are similar},
\item
replacement of an agent \emph{from the parent generation}.
\end{enumerate}
In Figure  \ref{fig:TwoPeaks.WF.AM10} the first three generations obtained by the iteration of this model are shown.
The only difference with respect to the pure random mating case (Figure  \ref{fig:TwoPeaks.WF}) is that the intermediate cluster does not appear because the interbreeding of a pair of individuals from either peak is prohibited by the assortativity condition.


\subsection{Cluster Formation in Opinion Dynamics}

From the point of view of self--organizing systems in opinion or cultural dynamics (e.g. \cite{Axelrod1997,Deffuant2001}) the result shown in Figure  \ref{fig:TwoPeaks.WF.AM10} is somewhat surprising because the introduction of interaction constraints is known to lead to co--existence of clusters of individuals (assortative mating is often called bounded confidence in this context).
This is even more interesting as the microscopic rules used to model the self--organization in opinion dynamics are very similar.
\begin{enumerate}
\item
selection of two individuals, \emph{all with equal probability}.
\item
application of recombination and mutation rules \emph{if the individuals are similar},
\item
\emph{update} of one parent agent.
\end{enumerate}

In this scheme, we emphasized differences with respect to the WF model.
Notice that mutations, sometimes interpreted as cultural drift, are not always taken into account.

Notice also that this form of replacement where effectively one parent individual is chosen to die to make place for the new--born is sometimes considered in population genetics (see, for instance, \cite{Moran1958,Korolev2010}).

In opinion dynamics the initial population is usually distributed according to the uniform distribution.
In general, there are no global influences such as a fitness landscape so that the probability of selection is equal for all individuals independent of their position in the trait space.


The locally--interacting model (henceforward called LI model) is implemented as a model of overlapping generations (OLG).
That is, the population is updated after each single interaction event (and not after $N$ events).
Notice that this means that the new state of an individual that is updated is from then on taken into account in the later iterations.
Therefore, a single iteration actually means a single interaction event involving two individuals.
Nevertheless, for the sake of comparability with the WF model, we can consider generations in the LI model by assuming that we pass from one generation to the next ($g\rightarrow g+1$) after $N$ iterations (interaction events).

\begin{figure*}[htp]
\centering
\begin{tabular}{ccc}
\includegraphics[width=0.28\linewidth]{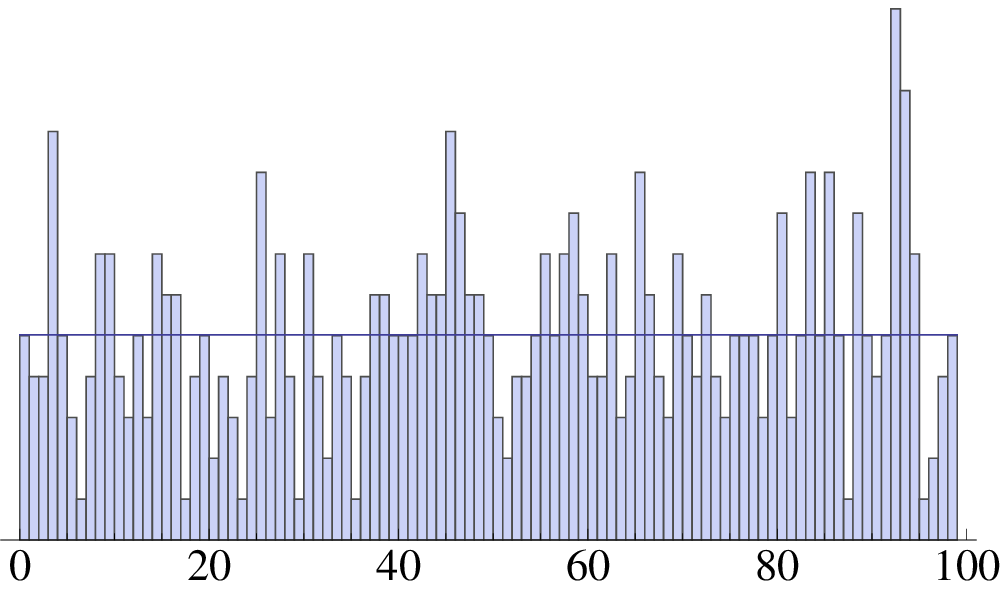}&
\includegraphics[width=0.28\linewidth]{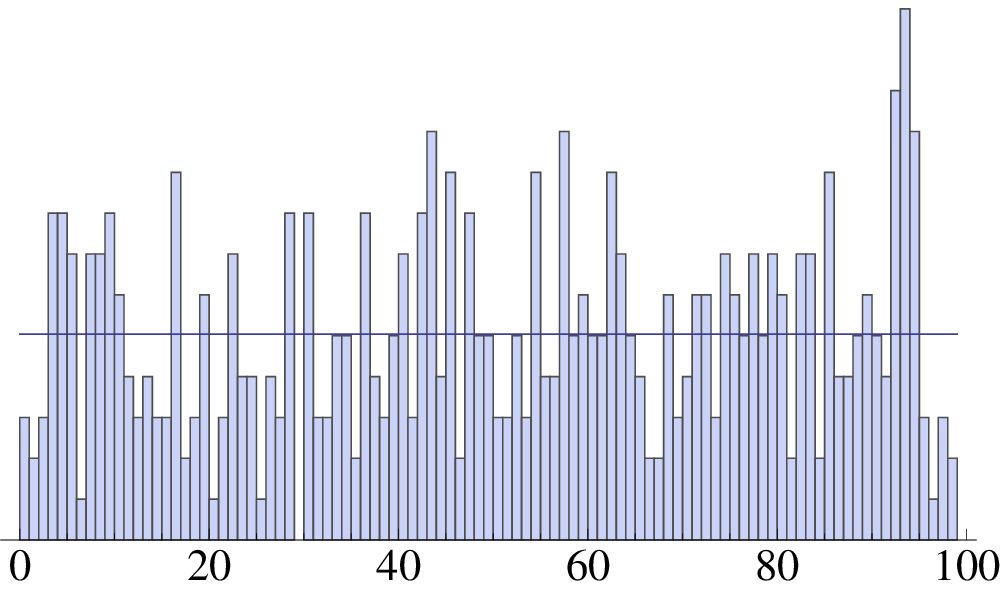}&
\includegraphics[width=0.28\linewidth]{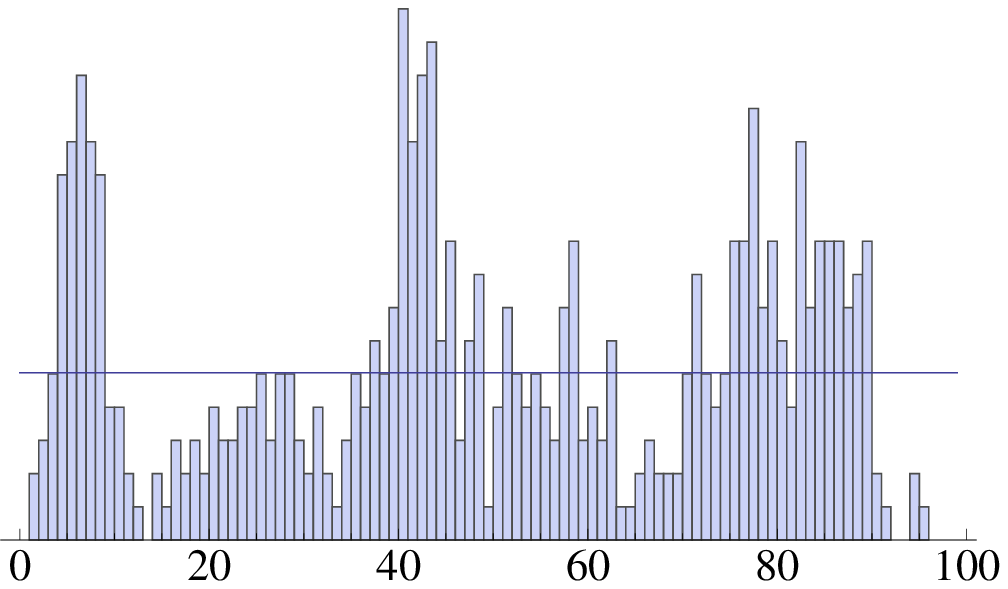}\\
\includegraphics[width=0.28\linewidth]{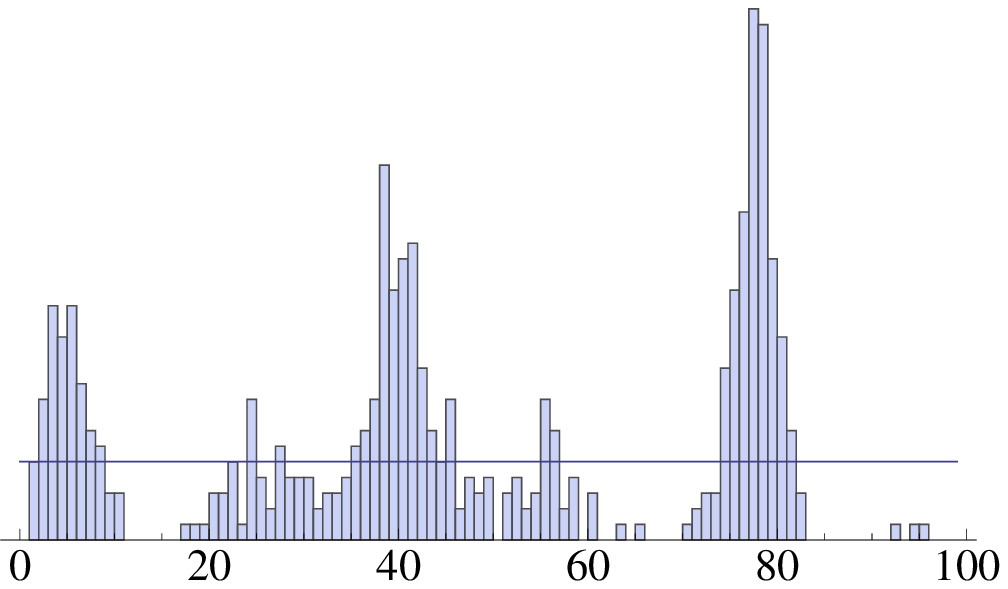}&
\includegraphics[width=0.28\linewidth]{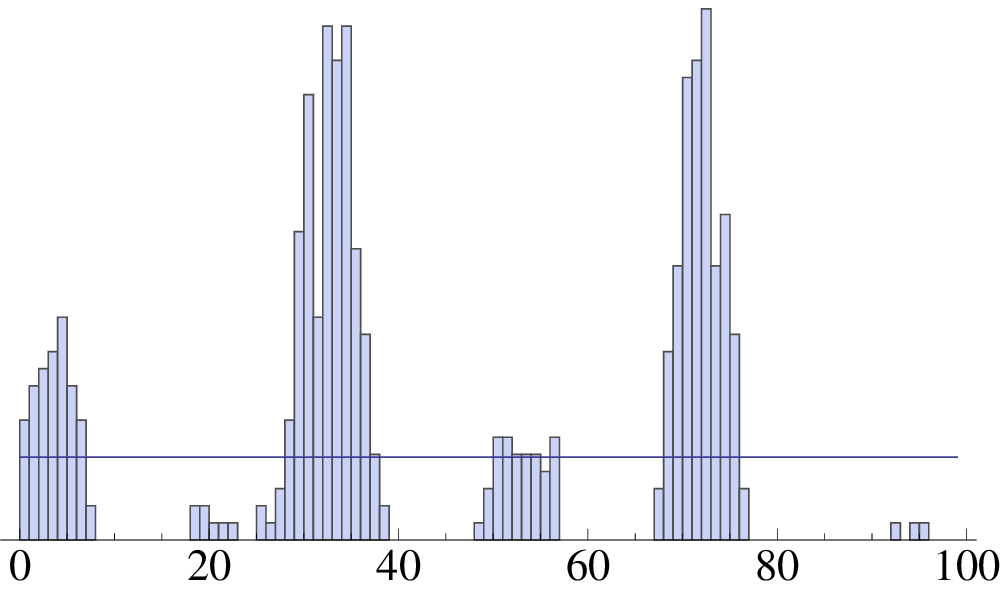}&
\includegraphics[width=0.28\linewidth]{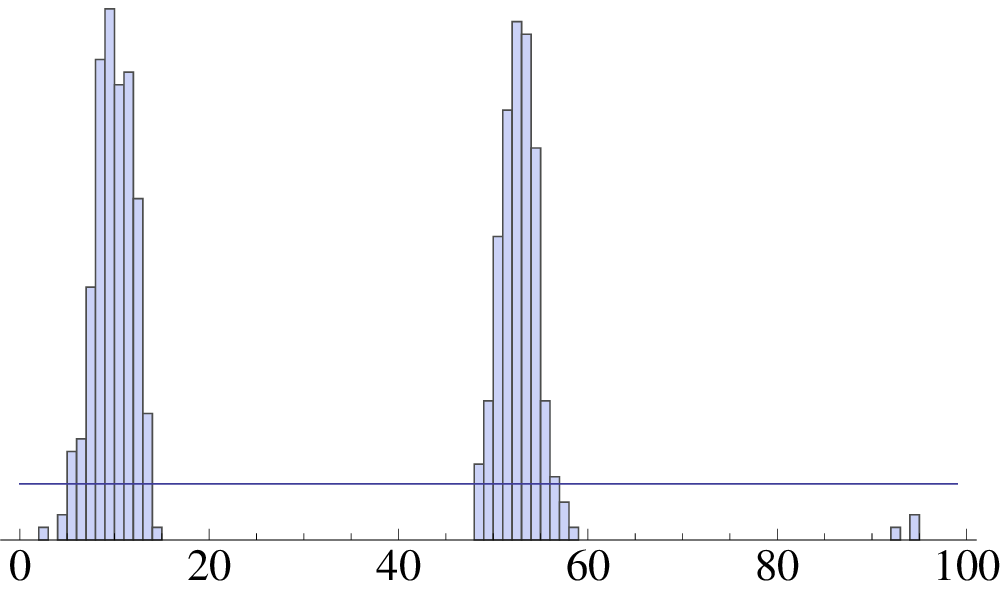}
\end{tabular}
\caption{Emergence of clustering in the LI model with assortativity. In the upper row the initial population and the distribution of the first and the 10th generation are shown from left to right ($g=0,1,10$). Bottom row shows, from left to right, the 20th, 40th and 100th generation ($g=20,40,100$).}
\label{fig:Flat.SO.AM10}
\end{figure*}

In Figure  \ref{fig:Flat.SO.AM10} we show a realization of the simulation for 500 individuals initially distributed uniformly over the traits from 0 to 99.
Update only takes place if the distance between two individuals is smaller then 10.
It becomes clear that initial inhomogeneities are reinforced during the process such that clusters of individuals are formed.
Compared to the WF simulations this process is slow.
In Figure  \ref{fig:Flat.SO.AM10} we show from top left to bottom right the original population ($g=0$), and the population in the 1st, 10th, 20th, 40th and 100th generation ($g = 1,10,20,40,100$).
From generations 40 to 100 some of the clusters have disappeared so that only two large sub--populations (and a very small one at around 90) remain.
In the long run these clusters might merge due to mutations (drift).
In any case the co--existence of "reproductively isolated" sub--populations is rather stable during long periods of the process.

\begin{figure*}[htp]
\centering
\begin{tabular}{ccc}
\includegraphics[width=0.28\linewidth]{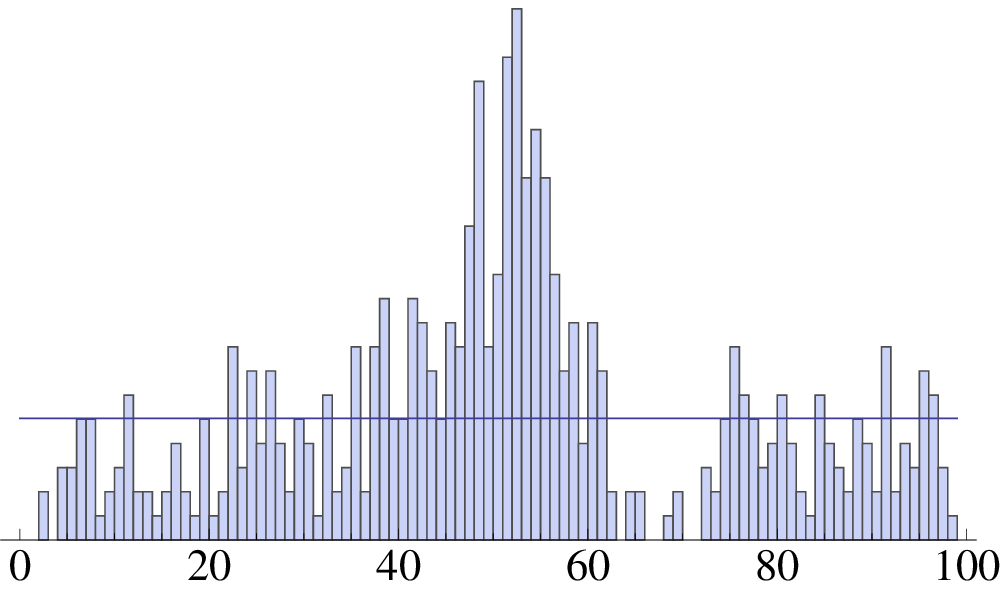}&
\includegraphics[width=0.28\linewidth]{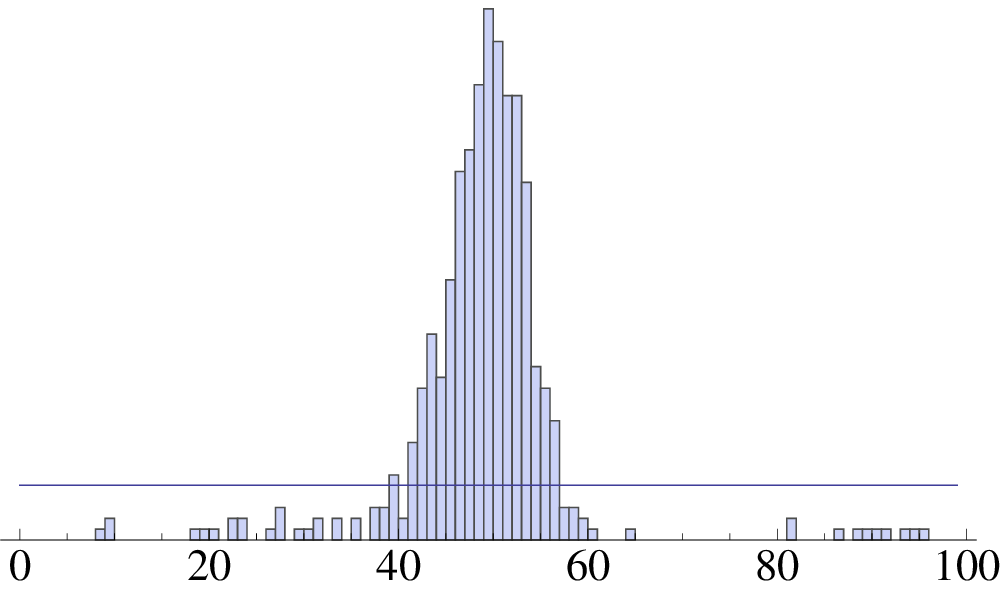}&
\includegraphics[width=0.28\linewidth]{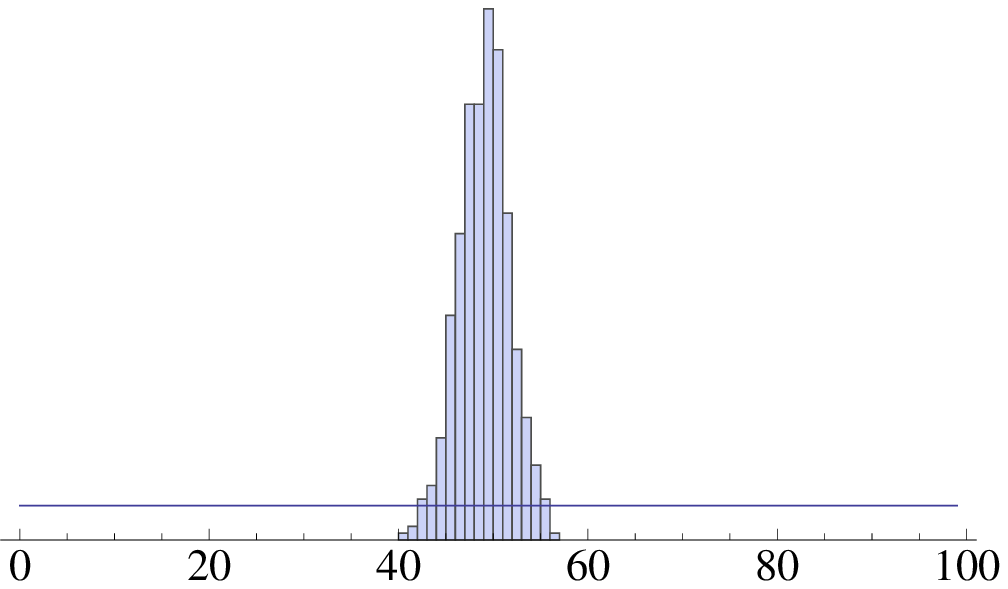}
\end{tabular}
\caption{WF model with overlapping generations with flat fitness landscape, assortative mating and uniform initial population as in Figure  \ref{fig:Flat.SO.AM10}. From left to right the first three generations ($g=1,2,3$) are shown.}
\label{fig:Flat.WF.AM10.OG}
\end{figure*}

\begin{figure*}[htp]
\centering
\begin{tabular}{ccc}
\includegraphics[width=0.28\linewidth]{OnePeak.WF.00.eps}&
\includegraphics[width=0.28\linewidth]{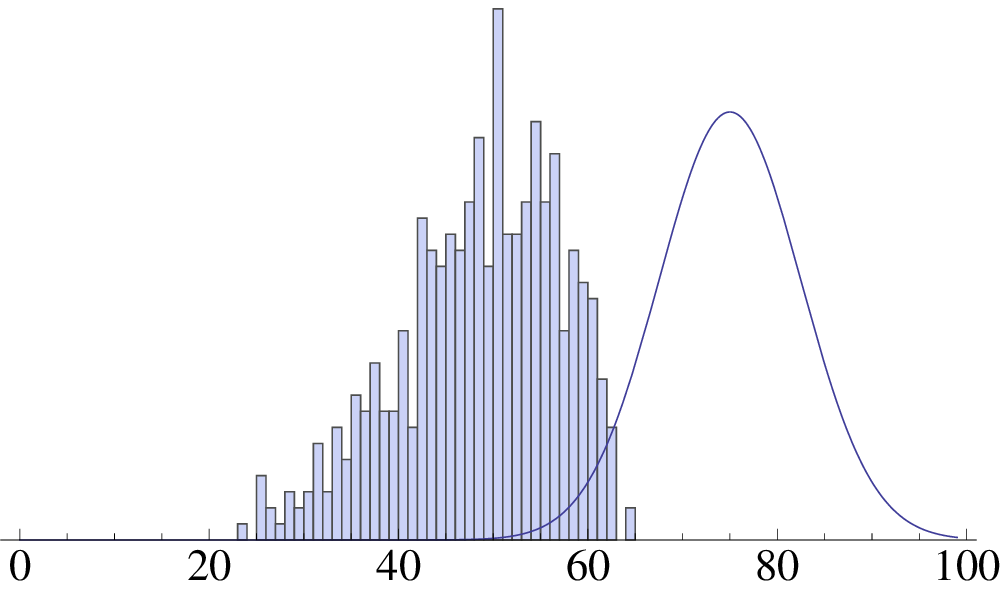}&
\includegraphics[width=0.28\linewidth]{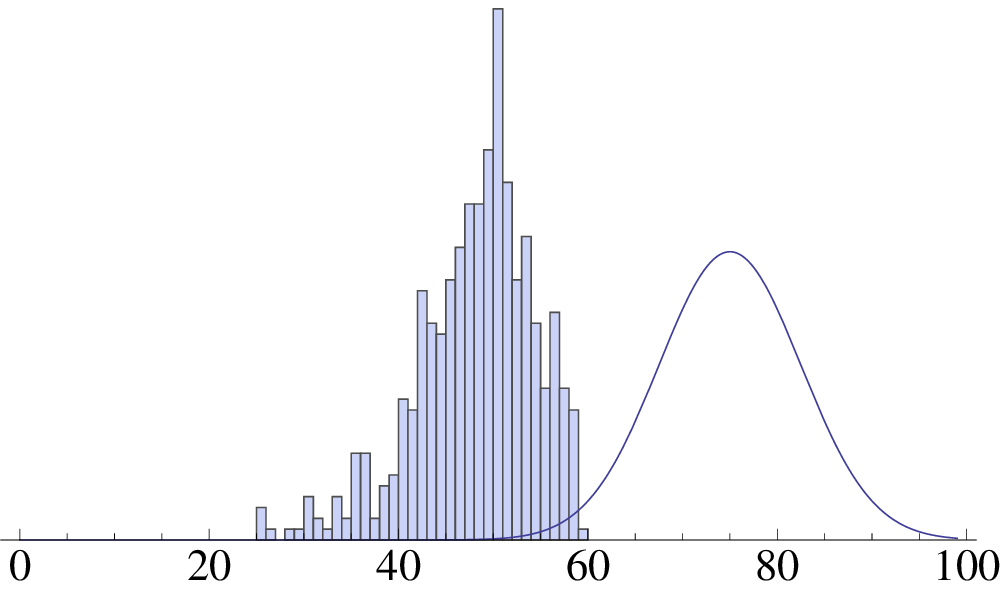}
\end{tabular}
\caption{The self--organization LI model with local update on a peaked fitness landscape. From left to right the initial population and the first two generations ($g = 0,1,2$) are shown.}
\label{fig:OnePeak.SO.AM40}
\end{figure*}

\subsection{Overlapping versus Non--overlapping Generations}

There seems to be a subtle difference between the LI model and the WF model, with a crucial effect, however.
There are three potential sources of the different behavior:
\begin{enumerate}
  \item  There is uniform fitness in the LI model but a peaked landscape in the WF model.
  \item  The LI model  is implemented as a model of OLG whereas the WF model implements NOLG.
  \item  In the LI model, the state change of an individual is modeled whereas the creation of a new individual is considered in the WF model.
\end{enumerate}

The first two cases can be checked easily by implementing the WF model with OLG and looking at a realization using the same conditions as in Figure  \ref{fig:Flat.SO.AM10}.
This is sometimes referred to as Moran model (\cite{Moran1958, Korolev2010}).
Two individuals give rise to a new individual which replaces another individual in the current generation.
The first three generations ($g = 1,2,3$ corresponding to the population after $500$, $1000$ and $1500$ sequential mating events) are shown in Figure  \ref{fig:Flat.WF.AM10.OG}.
The initial population is the same as before (upper left in Figure  \ref{fig:Flat.SO.AM10}).
The behavior of the model is in drastic contrast to the behavior of the LI scheme.
In fact, the behavior is very similar to the original WF model with NOLG.

We conclude that the qualitative differences between the WF model and the LI model are neither due to different ways of dealing with generations (OLG versus NOLG) nor to the choices of different fitness landscapes.


\subsection{Local versus Non--local Replacement}

It turns out that in the implementation of the WF model with OLG, a decision must be taken whether the new individual replaces one of its parents or an arbitrary individual from the generation and that the two alternatives result in qualitatively different dynamical behaviors.
We will call \emph{local replacement} the case that the new individual replaces one of its parents and \emph{non--local replacement} refers to the case that an arbitrary individual is replaced by the new one.
Noteworthy, there is a tendency that models with NOLG implement a form of non--local replacement because no care is usually taken about the order of individuals such that a child will in general appear at a position in the population array that is distant from the position of the parents.

In this way non--local replacement undermines the effects of assortative mating, because an individual with a trait $x$ can effectively be replaced by an individual with trait $y$ even if $|x - y| > 10$.



\subsection{(Non--)Adaptiveness of Local Replacement}

Earlier we saw that modeling speciation in a model with a fixed population size requires that the update process operates with local replacement.
However, it turns out that in this case the process looses its adaptiveness.
Figure  \ref{fig:OnePeak.SO.AM40} shows the first generations ($g=0,1,2$) of the self--organization model with local replacement performed on a fitness landscape with a single peak (compare Figure  \ref{fig:OnePeak.WF}).
The population is actually pushed away from the peak.
This is due to the fact that the individuals close to the peak, though frequently chosen, are not replaced by individuals with low fitness (rarely chosen) so that the proportion of fit individuals does not increase.
To the contrary, mutations tend to drive fittest individuals away from the peak.
Hence, the mode of replacement in these two models with almost the same microscopic rules has a dramatic effect on the dynamics behavior.
Cluster formation (or speciation) and adaptiveness are in the context of these models two opposing phenomena such that an explanation of the two together is not achieved by one and the same model.

\section{Probabilistic Analysis of a Minimal Model}
\label{sec:mathematic}

The simulations show that there are decisive differences between different implementations of the simulation models even though the microscopic rules of agent choice and recombination are in fact equal.
In particular, it turned out that the qualitative differences in the model behavior are due to different modes of agent replacement.
This section elaborates these differences for a minimal model where the number of allowed traits is reduced from 100 to three.
The model implements the same mechanisms as before, on this reduced space with three traits only, excepting mutations.
Looking at the rate (probabilities) of transitions from one trait to the other we derive Markov chains, and the transition structure of these chains inform us about the dynamic mechanisms which different replacement modes give rise to.

\subsection{A Minimal Model}

Consider that there are only three different phenetic traits: the states left ($L$), right ($R$) and intermediate ($M$).
As before, in every interaction event pairs of individuals are chosen and the state of the new individual is determined by the recombination of the parent states.
We do not consider mutations here.
In accordance with the recombination rule in the previous section, whenever two parents are in the same state, the child will also be in that state: we denote this by $LL \rightarrow L, MM \rightarrow M$ and $RR \rightarrow R$.
If one of the parents is in $L$ and the other in $R$ recombination will lead to $M$, that is, $LR \rightarrow M$ and $RL \rightarrow M$.
In case $L$ mates with $M$ we say that recombination leads to $M$ ($LM \rightarrow M$) and vice versa if $M$ mates with $L$ it leads to $L$ ($ML \rightarrow L$).
Likewise, for matings between $M$ and $R$--agents, we set $RM \rightarrow M$ and vice versa $MR \rightarrow R$.
Notice that in the case of complete, homogeneous mixing the choice probabilities are symmetric such that choosing two agents with $RM$ is equally likely as choosing them in reverse order $MR$.

Associated with each of these nine possible transitions we define an additional probability $\alpha$ to be the probability that the recombination step is indeed performed once the respective trait combination is chosen.
In the model without trait--dependent mating constraints or fitness differences, all the $\alpha$ are set to one.
The reason for introducing this probability is that we can model assortative mating by setting $\alpha_{LR} = \alpha_{RL} = 0$.
In that case, a pair of individuals in $L$ and $R$ are assumed to be unable to produce offsprings.
Because no state changes take place in that case, the respective probabilities now contribute to keeping whole population unchanged.

\subsection{Transition Rates}

\begin{figure*}[ht]
\centering
\begin{tabular}{ccc}
\includegraphics[width=0.35\linewidth]{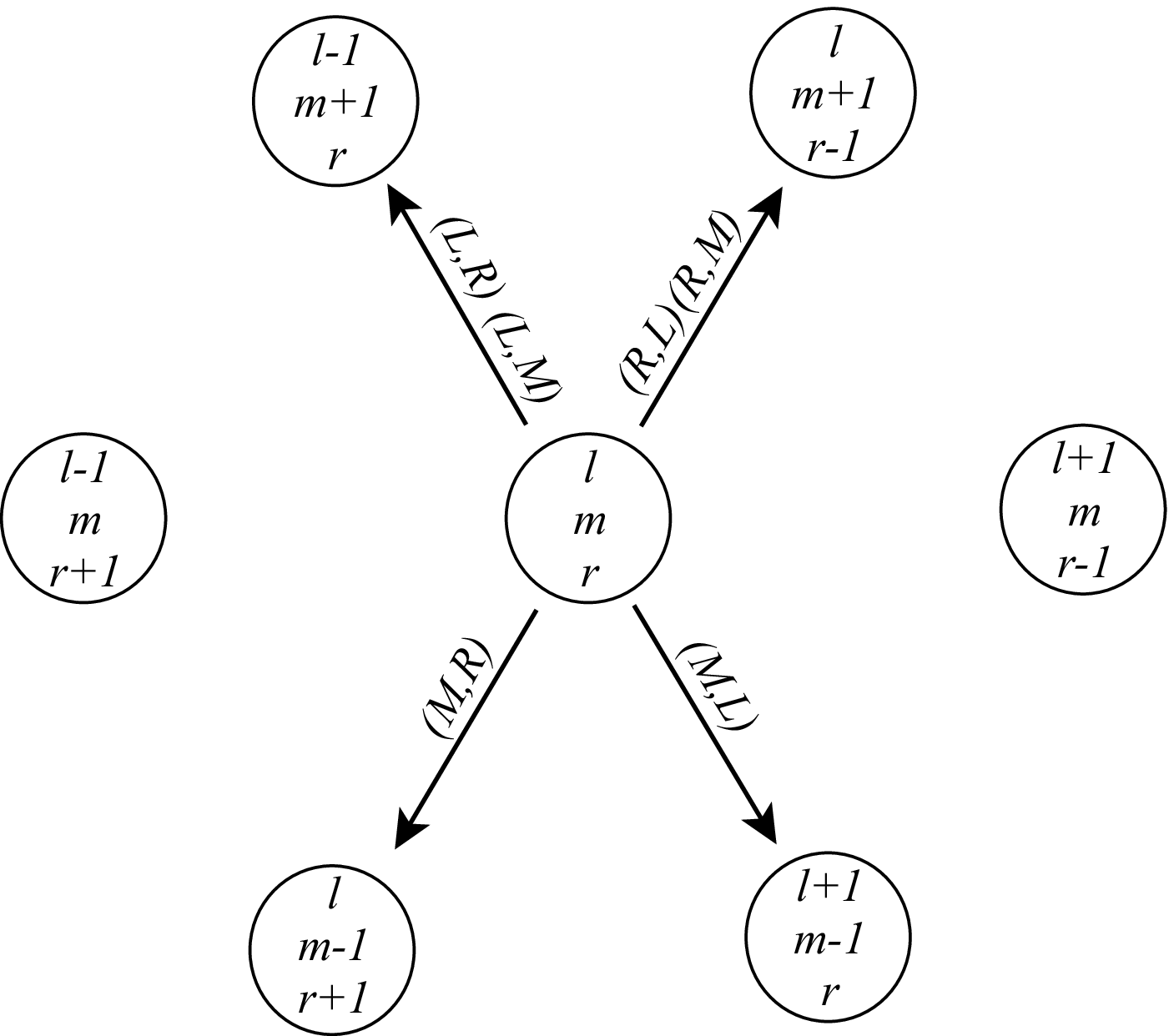} &
\hspace{64pt} &
\includegraphics[width=0.35\linewidth]{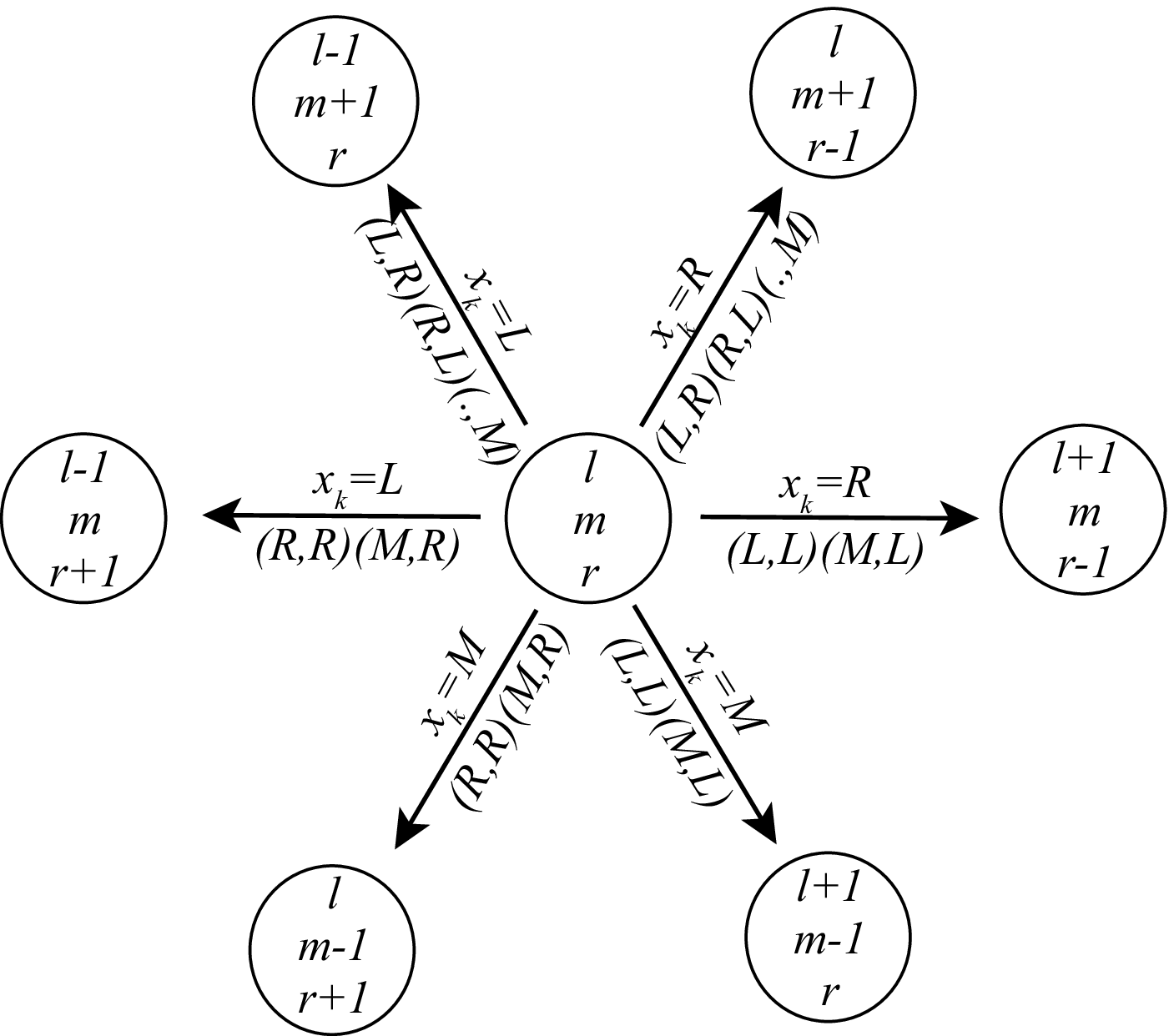}
\end{tabular}
\caption{Possible transitions in terms of the counters $l,m,r$ for local (l.h.s.) and random (r.h.s.) replacement.}
\label{fig:Transitions}
\end{figure*}

We consider a system of $N$ agents and characterize a population by counting the number of agents in the respective states $L, M$ and $R$.
Let us denote the number of agents in state $L$ by $l$, the number of $M$-agents by $m$ and the number of agent in $R$ by $r$.
After a mating event, the counters $l,m,r$ are either unchanged or one of them increases while another one decreases by one (ex. $l-1,m+1$).
The latter case simply means that one new individual (in state $M$) has replaced another one (in state $L$).
In the case of complete mixing, all the agents have equal probabilities to be chosen in the iteration process.
Therefore, we model the choice of an individual as a choice from an urn with $N$ balls of three different colors $L$, $M$ and $R$.
It is then clear that the choice of an individual with feature $F$ is $f/N$  (generic $F$).

In this way, it is possible to derive equations for the probabilities of all the possible changes of $l,m$ and $r$ from one mating event to the other.
Notice that already Moran adopted a similar Markov chain approach in the analysis of his model in \cite{Moran1958}.
A careful consideration of the relation between these macro--level equations and the microscopic simulation model is presented in \cite{Banisch2012}.

Let us denote the probability that $m$ increases while $l$ decreases by one as $P^m_l$.
For the model with local replacement we use the convention that the agent chosen first is always replaced by the new one.
Under this assumption the event $(l,m) \rightarrow (l-1,m+1)$ takes place if the states of the agent pair are either $(L,R)$ or $(L,M)$.
The probability that a pair $(L,R)$ is chosen is $\frac{lr}{N^2}$ which we denote as $p_{LR}$.\footnote{Notice that the agent choice is with replacement so that an individual may be chosen twice. This corresponds to self--fertilization and we allow it to keep the model as simple as possible.}
For $(L,M)$ we have $p_{LM} = \frac{lm}{N^2}$.
We integrate into this description the additional constraint $\alpha_{LR}$ ($\alpha_{LM}$) as the probability that the respective combination, once chosen, gives indeed rise to a new individual.
Then we obtain for probability $Pr[LR \rightarrow M] = \alpha_{LR} \ p_{LR}$ ($Pr[LM \rightarrow M] = \alpha_{LM} \ p_{LM}$).
With this definitions we obtain
\begin{equation}
P^m_l = \alpha_{LR}\frac{lr}{N^2}+\alpha_{LM}\frac{lm}{N^2} = \alpha_{LR} \ p_{LR}+\alpha_{LM} \ p_{LM}.
\label{eq:Plm.local}
\end{equation}
Equivalently, for the other non--zero transitions in the local case we find
\begin{eqnarray}
\begin{array}{l l}
P^m_r &= \alpha_{RL} \ p_{RL}+\alpha_{RM} \ p_{RM}\\
P^l_m &= \alpha_{ML} \ p_{ML}\\
P^r_m &= \alpha_{MR} \ p_{MR}.
\end{array}
\label{eq:P.local}
\end{eqnarray}

For the model with non-local (random) replacement we assume that the new--born individual replaces a randomly chosen agent.
The probability that this is an agent in state $F$ is again $f/N$ (generic $F$).
With this convention we find for the replacement of an $L$--agent
\begin{eqnarray}
\begin{array}{l l}
P^m_l &= \frac{l}{N}
(
\alpha_{LR} \ p_{LR} + \alpha_{RL} \ p_{RL} + \\
 & + \
\alpha_{LM} \ p_{LM} +
\alpha_{RM} \ p_{RM} +
\alpha_{MM} \ p_{MM} )\vspace{6pt}\\
P^r_l &= \frac{l}{N}
\left(
\alpha_{RR} \ p_{RR} + \alpha_{MR} \ p_{MR}
\right).
\end{array}
\label{eq:Pl.random}
\end{eqnarray}
For the replacement of an $R$--agent we have
\begin{eqnarray}
\begin{array}{l l}
P^m_r &= \frac{r}{N}
(
\alpha_{LR} \ p_{LR} + \alpha_{RL} \ p_{RL} +\\
 & + \
\alpha_{LM} \ p_{LM} +
\alpha_{RM} \ p_{RM} +
\alpha_{MM} \ p_{MM} )\vspace{6pt}\\
P^l_r &= \frac{r}{N}
\left(\alpha_{LL} \ p_{LL} + \alpha_{ML} \ p_{ML} \right),
\end{array}
\label{eq:Pr.random}
\end{eqnarray}
and for replacement of an $M$--agent
\begin{eqnarray}
\begin{array}{l l}
P^l_m &= \frac{m}{N}
\left(\alpha_{LL} \ p_{LL} + \alpha_{ML} \ p_{ML}
\right)\vspace{6pt}\\
P^r_m &= \frac{m}{N}\left(
\alpha_{RR} \ p_{RR} + \alpha_{MR} \ p_{MR}
\right).
\end{array}
\label{eq:Pm.random}
\end{eqnarray}
For a better orientation we visualize the possible transitions for both replacement modes along with the conditions for the transitions in Figure  \ref{fig:Transitions}. 


\begin{figure*}[ht]
\vspace{-0pt}
\centering
\begin{tabular}{ccc}
\includegraphics[width=0.48\linewidth]{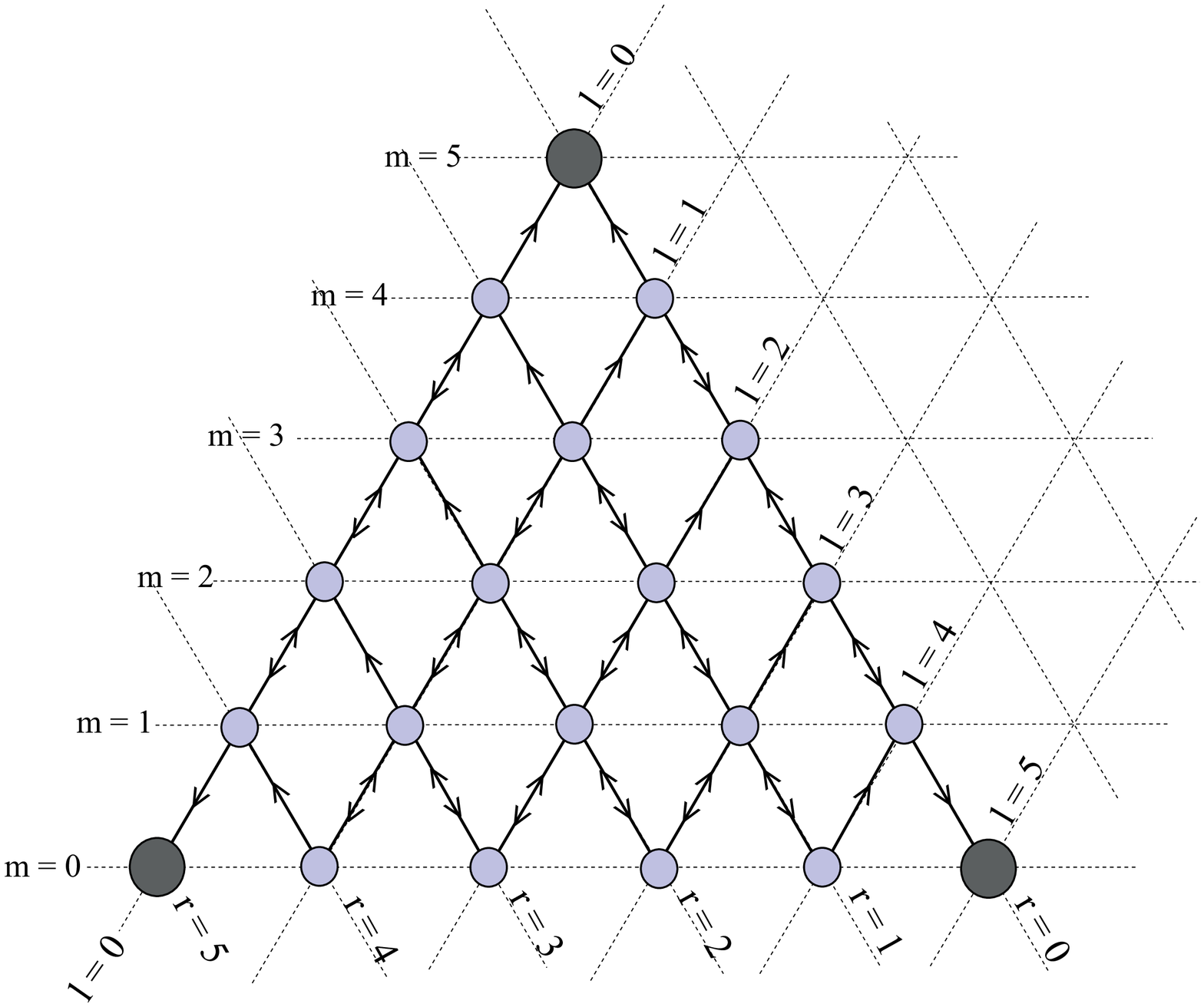}
&
\hspace{24pt}&
\includegraphics[width=0.48\linewidth]{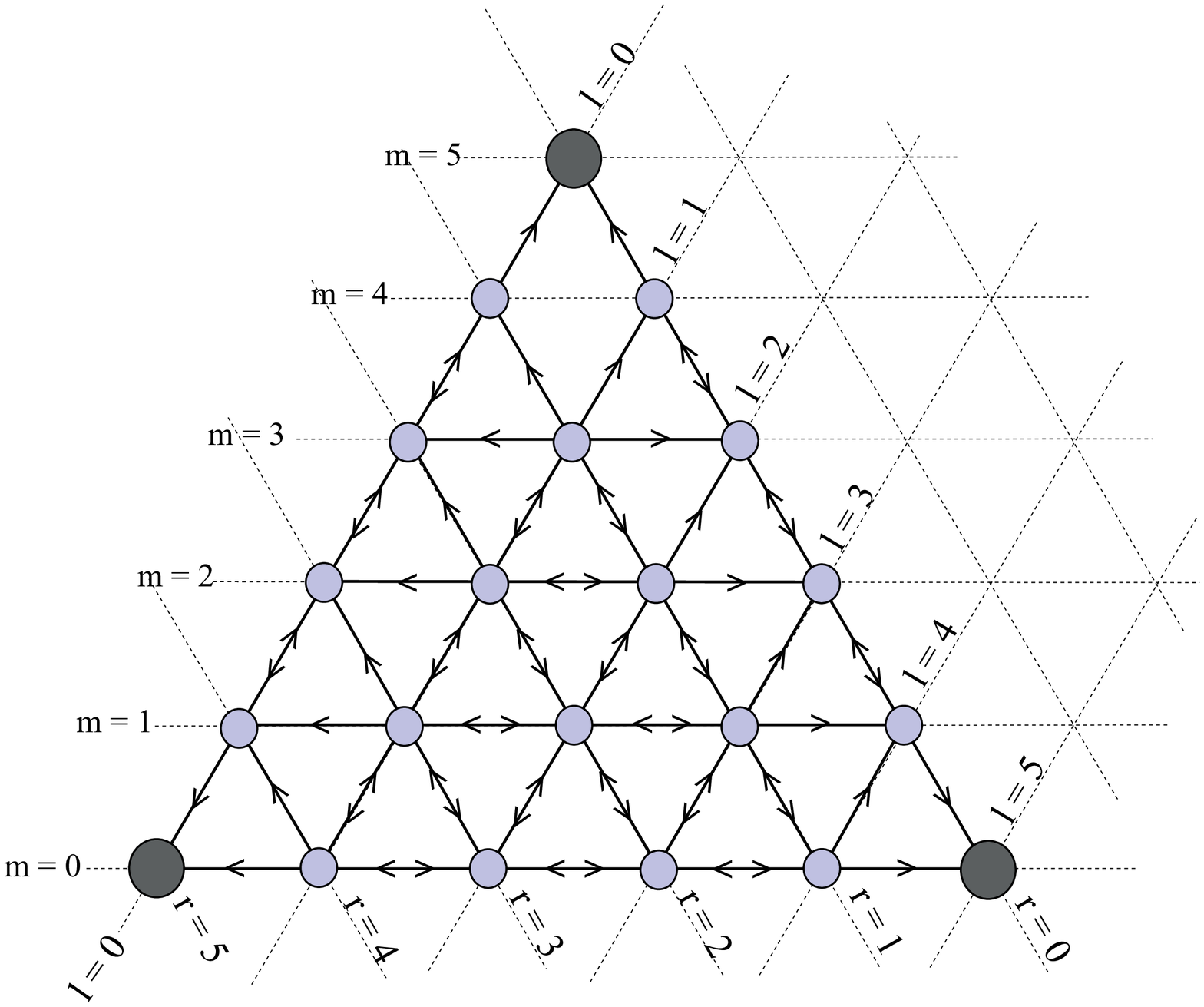}
\end{tabular}
\caption{Transition structure for local replacement (l.h.s.) and random replacement (r.h.s.) with random mating. In both cases, there are three absorbing states each corresponding to a homogeneous population.}
\label{fig:Chain.RandomMating}
\end{figure*}

\begin{figure*}[ht]
\centering
\begin{tabular}{ccc}
\includegraphics[width=0.48\linewidth]{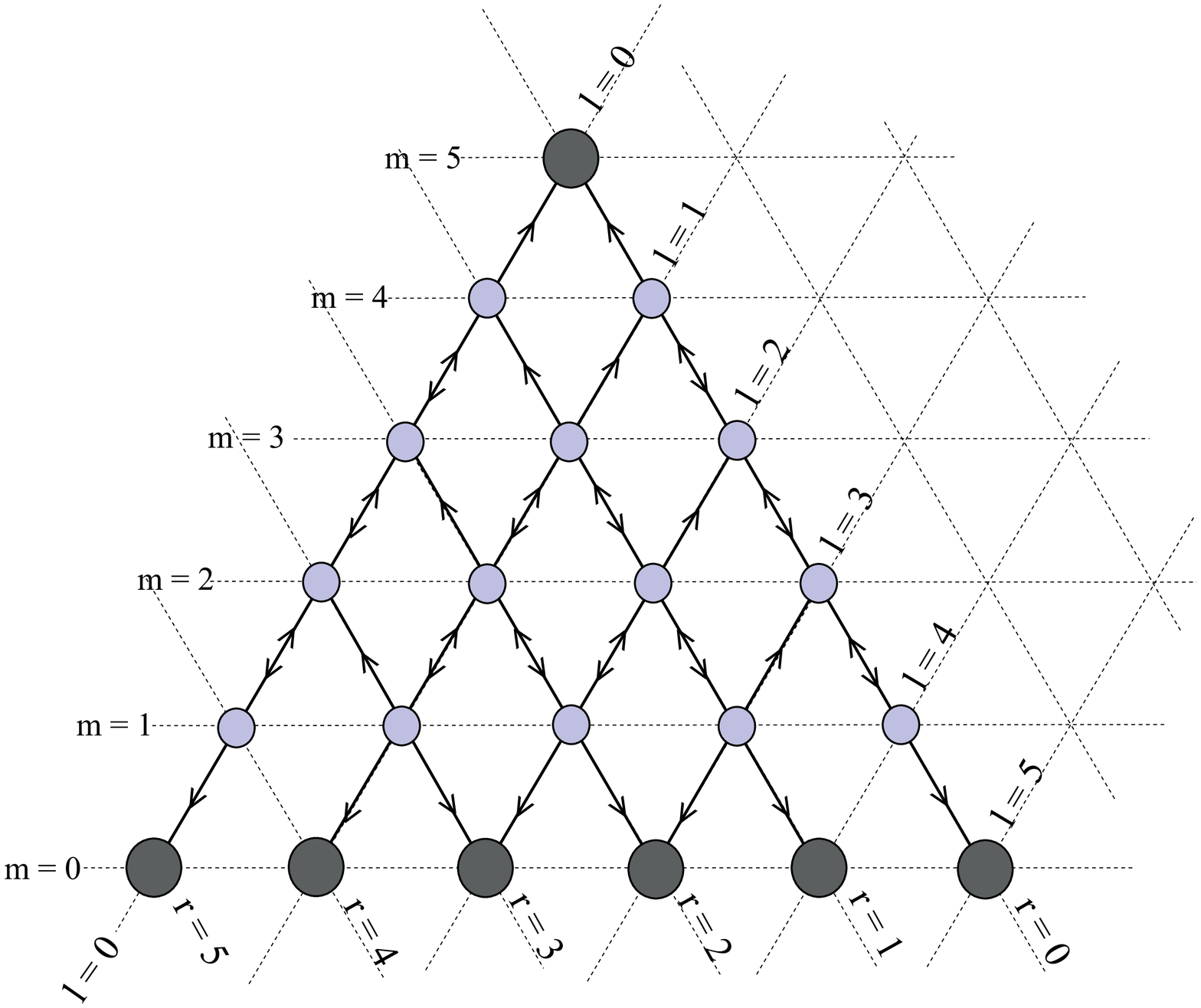}
&
\hspace{24pt}&
\includegraphics[width=0.48\linewidth]{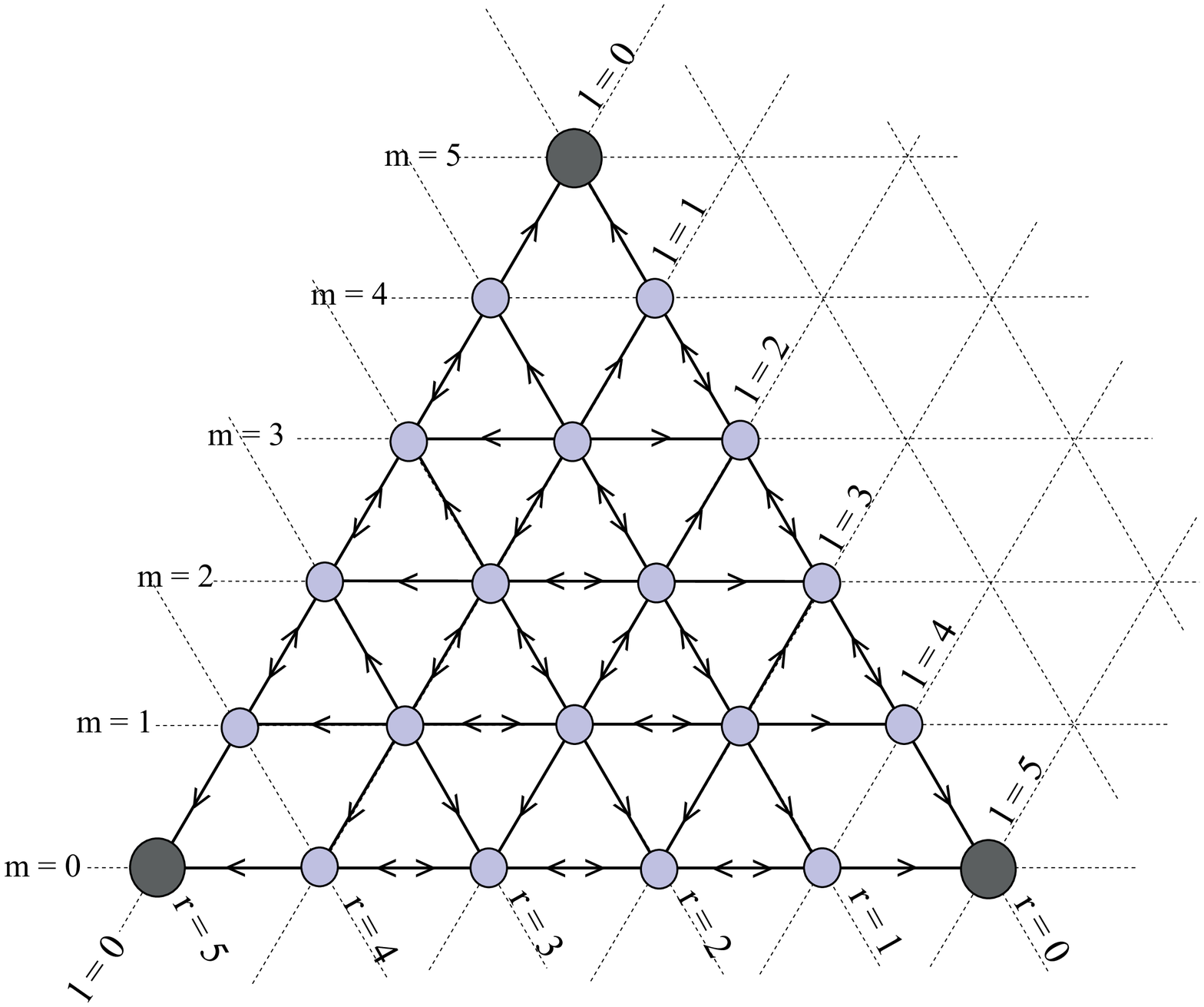}
\end{tabular}
\caption{Transition structure for local replacement (l.h.s.) and random replacement (r.h.s.) with assortative mating. Additional absorbing states emerge under local replacement, but not under random replacement as used in the WF model.}
\label{fig:Chain.AssortMating}
\end{figure*}

\subsection{Random Mating}

If all the $\alpha$ are equal to one, the transition equations (\ref{eq:Plm.local}) and (\ref{eq:P.local}) and respectively equations (\ref{eq:Pl.random}) to (\ref{eq:Pm.random}) realize all the transitions shown in Figure \ref{fig:Transitions} with a probability greater than zero.
In Figure \ref{fig:Chain.RandomMating} the complete transition structure is shown for the model with five agents.
Notice that for $N = 5$ each counter ($l,r,m$) can take values in between zero and five and that the triangular structure appears because we have $l+m+r = N$.

The larger gray atoms are the absorbing states of the process: they can be reached by a transition, but once reached, there is no transition leaving them.
Therefore they characterize the final configurations of the process.
For both local and non--local replacement the absorbing states are the three corners of the triangle grid with $l = N$ or $m = N$ or $r = N$.
This means that  the process will converge to a population with all individuals in the same state.
The smaller light--blue states indicate the transient atoms and the chances that the process remains in those atoms decreases exponentially with time (see, for instance, \cite{Seneta2006}).

\subsection{Assortative Mating}

The situation becomes different if we set $\alpha_{LR} = \alpha_{RL} =0$ by which we prohibit mating between $L$-- and $R$--agents.
This is assortative mating which means, in this simple model with only three traits, that left and right agents are incompatible and cannot produce offspring.
As noted above, the respective probabilities contribute to the probability that nothing changes as in that cases $(l,m,r) \rightarrow (l,m,r)$.
For both replacement modes the assortativity condition changes the transition probabilities and we compare the resulting transition structures in Figure  \ref{fig:Chain.AssortMating}.


Most importantly, for the local model, all the probabilities in (\ref{eq:Plm.local}) and (\ref{eq:P.local}) become zero if $m = 0$.
Hence, if there is no intermediate individual left ($m=0$) the process will remain where it is even if both $l$ and $r$ are larger than zero.
Assortative mating may therefore lead to the stable co--existence of $L$-- and $R$--agents.
Under non--local replacement this does not happen because even if certain transitions are canceled there remain horizontal transitions leading away from the respective two--species configurations.
This explains why speciation cannot be observed in the simulations performed in the first part of this paper.

It so happens that random replacement sets aside the effects of bounded confidence and consequently - like in the case of undirected genetic drift - leads to the merging of subpopulations.
As random interbreeding contributes to conservative dynamics, random replacement is also an opposing force to speciation.
This is due to the fact that under this replacement mode, a newcomer agent may take the place of a former-distant one. In so doing, forbidden transitions turn out to be allowed so that the consequences at the macro level become the same of unbounded confidence.

\subsection{Two--Peaked Fitness Landscape}

Next, let us discuss an extreme case of a two--peaked fitness function.
We consider the case that intermediate individuals have a zero fitness which we model by prohibiting all matings in which $M$--agents are involved.
This situation can be obtained  by assigning a zero probability to all the respective transitions, that is: $\alpha_{LM} = \alpha_{ML} = \alpha_{MM}=\alpha_{RM} = \alpha_{MR} = 0$.

From equation (\ref{eq:P.local}) we see that in the case of local replacement this leads to the strange situation that the probabilities for all those transitions by which the number of intermediates decreases become zero: $P^l_m = P^r_m = 0$.
Unless initialized with all agents in $L$ or $R$, the simulation will converge to the situation where all individuals are in the intermediate state ($M$), with zero fitness.
This clearly points at a deficiency of modeling adaptive dynamics with local replacement.
\\


All in all, we can conclude that adaptiveness is favored by non--local replacement while it is difficult to achieve speciation.
As opposed to this, under local replacement speciation becomes a natural result of assortative mating, but then the process is not convenient for approaching adaptive peaks in a fitness landscape.

\section{Discussion and Concluding Remarks}

In the context of the models we study in this paper, evolution by natural selection and locally interacting dynamics do not appear as opposing one another. In fact, the dynamical update rules used in the modeling of the microscopic interactions follow the same principles.

Let us try to adopt a broader perspective and to figure out a general framework comprising the main mechanisms leading to the emergence of collective structures in adaptive and self-organizing complex systems.

Back to the two phenomena underlying the paper research question,
we may say that the main consequences to the >>modelability<<  of either adaptation or speciation
are due to the constraints imposed on each of the above mechanisms of selection, interaction and replacement.
Their interplay is summarized in Table \ref{tab:1}.

\begin{table*}[ht]
\begin{center}
\begin{tabular}{c c c}
\hspace{0cm} \emph{Mechanisms}  & \hspace{5.0cm} & \emph{Emergent Patterns}
\end{tabular}
\begin{tabular}{|c|c|c||c|c|}
\hline
 Selection & Interaction   &Replacement& Outcome    & Example \\ \hline
     1 peak  & random      &  random   & \emph{convergence with Adaptation} & Figure 3 \\ \hline
     2 peaks & random      &  random   & \emph{convergence with Adaptation} & Figure 4 \\ \hline
     1 peak  & Assortative &  random   & \emph{convergence with Adaptation} & Figure 5 \\ \hline
     random  & Assortative &  Local    & \emph{speciation}    & F.6 and F.11(a)\\ \hline
     random  & Assortative &  random   & \emph{convergence} & F.7 and F.11(b)\\ \hline
     1 peak  & Assortative &  Local    & \emph{convergence without Adaptation}    & Figure  8 \\ \hline
     2 peaks & Assortative &  Local    & \emph{convergence without Adaptation}    & --- (Sect.3.5) \\ \hline
      random & random      &  Local    & \emph{convergence} & Figure 10(a) \\ \hline
      random & random      & random    & \emph{convergence} & Figure 10(b) \\ \hline
\end{tabular}
\medskip
\caption{General Framework}
\label{tab:1}
\end{center}
\end{table*}

The framework presented in Table 1 schematically shows the consequences of adopting (un)constrained mechanisms to the emergent outcome of a SO process.
It helps to emphasize that the emergence of some specific patterns may be strongly dependent on the way constraints dictate limitations on the selection,
interaction and replacement mechanisms. More specifically, it shows that differently (un)constraining the replacement mechanism of an SO process provides the conditions required for either speciation (the emergence of multi-modal distributions) or adaptation, since these features appear as two opposing phenomena,
not achieved by one and the same model.

In the same way that random interbreeding leads to conservative dynamics, random replacement is also
an opposing force to speciation since newcomers may take the place of former-distant agents.
In so doing, at the macro level, random replacement sets aside the effect of bounded confidence and -
like undirected genetic drift - may lead to the merging of subpopulations.

Even though we show in this paper that natural selection, operating as an external, environmental mechanism, is neither necessary nor sufficient for the creation of clustered populations, we do not want to argue against natural selection as an important mechanism in the biological domain and a substantive driving force in the speciation process.
To the contrary, the concept of (natural) selection operating at a global level may provide us with plausible interpretations of the model results, even in disciplines where such interpretations are still lacking.
In the words of T. Dobzhanski (\cite{Dobzhanski1970}, p.5-6):
\begin{quote}
[...] in biology nothing makes sense except in the light of evolution.
It is possible to describe living beings without asking questions about their origins.
The descriptions acquire meaning and coherence, however, only when viewed in the perspective of evolutionary development.
\end{quote}

\section*{Acknowledgement}

Financial support of the German Federal Ministry of Education and Research (BMBF) through the project \emph{Linguistic Networks} is gratefully acknowledged ({\tt http://project.linguistic-networks.net}).
This work has also benefited from financial support from the Funda\c{c}\~{a}o  para a Ci\^{e}ncia e a Tecnologia (FCT), under
the \emph{13 Multi-annual Funding Project of UECE, ISEG, Technical University of Lisbon}.

\small

\end{document}